\documentclass[pra,aps,epsfig]{revtex4}
\newcommand{\be}{\begin{equation}}
\newcommand{\ee}{\end{equation}}
\newcommand{\br}{\begin{eqnarray}}
\newcommand{\er}{\end{eqnarray}}
\newcommand{\nn}{\nonumber}
\newcommand{\bd}{\begin{displaymath}}
\newcommand{\ed}{\end{displaymath}}
\newcommand{\bfig}{\begin{figure}}
\newcommand{\efig}{\end{figure}}
\usepackage{graphicx}
\usepackage{amsfonts,amssymb,amsxtra,amsgen,amsopn}
\usepackage{amsmath,amsthm,amsbsy,amscd,amstext}
\def\alf{\alpha}

\def\om{\omega}
\def\eps{\epsilon}
\def\rpar{\right)}
\def\lpar{\left(}
\def\rbk{\right]}
\def\lbk{\left[}
\def\rbr{\right\}}
\def\lbr{\left\{}
\def\lb{\label}

\def\im{{\rm i}}
\def\tr{\mbox{${\rm Tr}$}}

\def\ro{\mbox{\boldmath $\rho$}}
\def\sig{\mbox{\boldmath $\sigma$}}
\def\bet{\mbox{\boldmath $\beta$}}
\def\vfi{\mbox{\boldmath $\varphi$}}
\def\bu{\mbox{\boldmath ${\cal U}$}}

\def\ima{\mbox{${\rm Im}$}}
\def\re{\mbox{${\rm Re}$}}
\def\rg{\rangle}
\def\lg{\langle}
\def\half{\frac{1}{2}}
\begin{document}
%
\title{Qualitative aspects of entanglement in the Jaynes-Cummings model with an external quantum field}
\author{Marcelo A. Marchiolli$^{1}$, Ricardo J. Missori$^{2}$ and
  Jos\'{e} A. Roversi$^{2}$} 
\affiliation{$^{1}$ Instituto de F\'{\i}sica de S\~{a}o Carlos, Universidade de S\~{a}o Paulo, \\
             Caixa Postal 369, 13560-970 S\~{a}o Carlos, SP, Brazil, \\
             Eletronic address: marcelo$\_$march@bol.com.br\\
             $^{2}$Instituto de F\'{\i}sica ``Gleb Wataghin", Universidade Estadual de Campinas, \\
             13083-970 Campinas, S\~{a}o Paulo, Brazil, \\
             Eletronic address: missori@ifi.unicamp.br and roversi@ifi.unicamp.br}
\date{\today}
%
\begin{abstract}
\vspace*{0.1mm}
\begin{center}
\rule[0.1in]{142mm}{0.4mm}
\end{center}
We present a mathematical procedure which leads us to obtain analytical solutions for the atomic inversion and Wigner 
function in the framework of the Jaynes-Cummings model with an external quantum field, for any kinds of cavity and driving fields. Such solutions
are expressed in the integral form, with their integrands having a commom term that describes the product of the 
Glauber-Sudarshan quasiprobability distribution functions for each field, and a kernel responsible for the entanglement. 
Considering two specific initial states of the tripartite system, the formalism is then applied to calculate the atomic 
inversion and Wigner function where, in particular, we show how the detuning and amplitude of the driving field modify the 
entanglement. In addition, we also obtain the correct
quantum-mechanical marginal distributions in phase space.
(Published in J. Phys. A: Math. Gen. {\bf 36}, 12275 (2003)) \\ 
\vspace*{0.1mm}
\begin{center}
\rule[0.1in]{142mm}{0.4mm}
\end{center}
\end{abstract}
\maketitle
\section{Introduction}

The concept of entanglement naturally appears in quantum mechanics when the superposition principle is applied to composite 
systems. In this sense, a multipartite system is entangled when their physical properties cannot be described through a 
tensor product of density operators associated to their different parts which constitute the whole system. An immediate 
consequence of this important effect has its origin in theory of quantum measurement {ref1}: the entangled state of 
the multipartite system can reveal information on its constituent parts. However, this information is extremely sensitive 
to the dissipative coupling between the macroscopic meter and its environment. In fact, entangled states involving 
macroscopic meters are rapidly transformed into statistical mixtures of product states and this fast relaxation process 
characterizes the decoherence \cite{ref2,ref3,ref4}. According to Raimond et al. \cite{ref5}: ``The decoherence itself 
involves entanglement since the meter gets entangled with its environment. As the information leaks into the environment, 
the meter's state is obtained by tracing over the environment variables, leading to the final statistical mixture. This 
analysis is fully consistent with the Copenhagen description of a measurement". Beyond these fundamental features, entangled
states have potential applications for information processing and quantum computing \cite{ref6,ref7,ref8,ref9}, quantum
teleportation \cite{ref10}, dense coding \cite{ref11}, and quantum cryptographic schemes \cite{ref12}.

A feasible physical system to generate entangled states is given by the Jaynes-Cummings model (JCM) which describes the
matter-field interaction \cite{ref13,ref14}. It is typically realized in cavity QED experiments involving Rydberg atoms
crossing superconducting cavities (one by one) in different frequency regimes and configurations, with relaxation rates
small and well understood \cite{ref5}. Recently, many authors have investigated the two-mode and driven JCM in different
contexts and predicted new interesting results \cite{ref15,ref16,ref17,ref18,ref19,ref20,ref21,ref22,ref23,ref24,ref25,
ref26,ref27}. For instance, Solano et al. \cite{ref26} have proposed a method of generating multipartite entanglement
through the interaction of a system of $N$ two-level atoms in a cavity of high quality factor with a strong classical
driving field. Following the authors, the main advantage of this external field in the system under consideration is the
great flexibility in generating entangled states, since it provides freedom in choosing the detuning and strength of the
field. On the other hand, Wildfeuer and Schiller \cite{ref27} have used the Schwinger's oscillator model to obtain a
mathematical solution for the generation of entangled $N$-photon states in the framework of the two-mode JCM. Here we
develop a mathematical procedure which permits us to obtain compact solutions for atomic inversion and Wigner function
in the framework of the driven JCM, considering any cavity and external fields. In particular, both solutions are expressed
in the integral form with their integrands presenting a common term that describes the product of the Glauber-Sudarshan
quasiprobability distributions \cite{ref28} for each field, and a kernel responsible for the correlations. To illustrate
our results we fix the cavity field in the even- and odd-coherent states \cite{ref29}, and the driving field in the
coherent state. Furthermore, we show how the detuning and amplitude of the driving field modify the entanglement in the
tripartite system via Wigner function.

The paper is organized as follows. In Section II we obtain the time-evolution operator and the matrix elements of the
density operator for the driven JCM, with the cavity and external fields described in the diagonal representation of
coherent states. Following, we fix the cavity field in the even- and odd-coherent states and the driving field in the
coherent state to investigate, in Section III, the effects of amplitude of the driving field and detuning parameters upon
the atomic inversion. In Section IV we derive a formal expression for the Wigner function associated with the cavity field
which permits us to analyse how the entanglement is modified in the tripartite system. Moreover, we also obtain analytical 
expressions for the correct quantum-mechanical marginal distributions in phase space. Section V contains our summary and 
conclusions. Finally, Appendixes A and B contain the main steps to calculate the atomic inversion and Wigner function, 
respectively.

\section{Algebraic aspects of the JCM with an external quantum field}

In general, the driven JCM consists of a two-level atom interacting nonresonantly with a single-mode cavity field, and
driven additionally by an external field through one open side of the cavity (the experimental scheme in the context of
cavity QED is sketched in figure 1). Within the dipole and rotating-wave approximations, the dynamics of the atom-cavity
system is governed by the Hamiltonian ${\bf H} = {\bf H}_{0} + {\bf V}$, where
\br
\lb{e1}
{\bf H}_{0} &=& \hbar \om \lpar {\bf a}^{\dagger} {\bf a} + {\bf b}^{\dagger} {\bf b} \rpar + \half \hbar \om \,
\sig_{{\rm z}}
\; , \\
\lb{e2}
{\bf V} &=& \half \hbar \delta \, \sig_{{\rm z}} + \hbar \kappa_{{\rm a}} \lpar {\bf a}^{\dagger} \sig_{-} + {\bf a}
\sig_{+} \rpar + \hbar \kappa_{{\rm b}} \lpar {\bf b}^{\dagger} \sig_{-} + {\bf b} \sig_{+} \rpar \; .
\er
Here, $\om$ is the cavity field frequency (we assume the resonance condition between the cavity and driving fields),
$\om_{0}$ is the atomic transition frequency, $\delta = \om_{0} - \om$ is the detuning frequency, and $\kappa_{{\rm a}
({\rm b})}$ is the coupling constant between the atom and the cavity (external) field. The atomic spin-flip operators
$\sig_{\pm}$ and $\sig_{{\rm z}}$ are defined as $\sig_{+} = | e \rg \lg g |$, $\sig_{-} = | g \rg \lg e |$, and
$\sig_{{\rm z}} = | e \rg \lg e | - | g \rg \lg g |$ ($| g \rg$ and $| e \rg$ correspond to ground and excited states of
the atom), with the following commutation relations: $\lbk \sig_{{\rm z}},\sig_{\pm} \rbk = \pm 2 \sig_{\pm}$ and $\lbk
\sig_{+},\sig_{-} \rbk = \sig_{{\rm z}}$. Furthermore, ${\bf a}$ $\lpar {\bf a}^{\dagger} \rpar$ and ${\bf b}$ $\lpar
{\bf b}^{\dagger} \rpar$ are the annihilation (creation) operators of the single-mode cavity and external fields,
respectively. It is important to mention that the quantum nature of the fields used in many
proposed schemes for quantum information processing present serious consequences in large
scale quantum computations, since the uncertainty principle and the possibility of becoming
entangled with the physical qubits represent possible limitations on quantum computing
\cite{ref30,ref31,ref32,ref33,ref34}. In this sense, van Enk and Kimble \cite{ref31} have considered the interaction of atomic
qubits laser fields and quantify atom-field entanglement in various situations of interest where,
in particular, they found that the entanglement decreases with the mean number of photons $\lg{\bf n}\rg$ in a laser beam as $E\propto \lg {\bf n} \rg ^{-1}
{\rm log}_{2}\lg{\bf n}\rg$ for $\lg {\bf n} \rg \gg 1$. Pursuing this line, Gea-Banacloche \cite{ref32}
has investigated the quantum nature of the laser fields used in the manipulation of quantum
information, focusing especially on phase errors and their effects on error-correction schemes
(for more details, see \cite{ref33,ref34}).
\begin{figure}[!t]
\centering
\begin{minipage}[b]{0.30\linewidth}
\includegraphics[width=\linewidth, angle=-90]{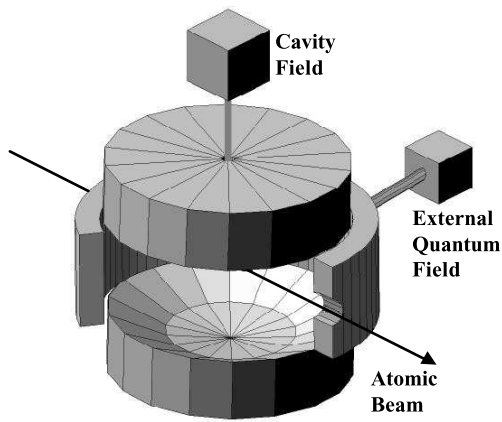}
\end{minipage}
\caption{Experimental apparatus used in the description of the JCM with an external quantum field.}
\end{figure}

Now, let us define the quasi-mode operators ${\bf A} = \eps_{{\rm a}} {\bf a} + \eps_{{\rm b}} {\bf b}$ and ${\bf B} =
\eps_{{\rm b}} {\bf a} - \eps_{{\rm a}} {\bf b}$ (where $\eps_{{\rm a}({\rm b})} = \kappa_{{\rm a}({\rm b})} /
\kappa_{{\rm eff}}$, and $\kappa_{{\rm eff}}^{2} = \kappa_{{\rm a}}^{2} + \kappa_{{\rm b}}^{2}$ is an effective coupling
constant), which satisfy the commutation relations
\br
\lbk {\bf A},{\bf A}^{\dagger} \rbk &=& {\bf 1} \; , \qquad \lbk {\bf N}_{{\rm A}},{\bf A} \rbk = - {\bf A} \; , \qquad
\lbk {\bf N}_{{\rm A}},{\bf A}^{\dagger} \rbk = {\bf A}^{\dagger} \; , \nn \\
\lbk {\bf B},{\bf B}^{\dagger} \rbk &=& {\bf 1} \; , \qquad \lbk {\bf N}_{{\rm B}},{\bf B} \rbk = - {\bf B} \; , \qquad
\lbk {\bf N}_{{\rm B}},{\bf B}^{\dagger} \rbk = {\bf B}^{\dagger} \; , \nn \\
\lbk {\bf A},{\bf B} \rbk &=& 0 \; , \qquad \lbk {\bf A},{\bf B}^{\dagger} \rbk = 0 \; , \qquad \lbk {\bf N}_{{\rm A}},
{\bf N}_{{\rm B}} \rbk = 0 \; , \nn
\er
being ${\bf N}_{{\rm A}} = {\bf A}^{\dagger} {\bf A}$ $\lpar {\bf N}_{{\rm B}} = {\bf B}^{\dagger} {\bf B} \rpar$ the
number operator related to the quasi-mode operator ${\bf A}$ $({\bf B})$. Introducing the number-sum operator ${\bf S} =
{\bf N}_{{\rm A}} + {\bf N}_{{\rm B}}$ and the number-difference ${\bf D} = {\bf N}_{{\rm A}} - {\bf N}_{{\rm B}}$, we
verify that: \\
(i) ${\bf S} = {\bf n}_{{\rm a}} + {\bf n}_{{\rm b}}$ is a conserved quantity (${\bf n}_{{\rm a}} = {\bf a}^{\dagger}
{\bf a}$ and ${\bf n}_{{\rm b}} = {\bf b}^{\dagger} {\bf b}$ are the photon-number operators of the cavity and external
fields); \\
(ii) the operator ${\bf D}$ can be written in terms of the generators $\{ {\bf K}_{+},{\bf K}_{-},{\bf K}_{0} \}$ of the
${\rm SU}(2)$ Lie algebra,
\be
\lb{e3}
{\bf D} = 2 \lpar \eps_{{\rm a}}^{2} - \eps_{{\rm b}}^{2} \rpar {\bf K}_{0} + 2 \eps_{{\rm a}} \eps_{{\rm b}} \lpar
{\bf K}_{-} + {\bf K}_{+} \rpar \; ,
\ee
where ${\bf K}_{-} = {\bf a} {\bf b}^{\dagger}$, ${\bf K}_{+} = {\bf a}^{\dagger} {\bf b}$, and ${\bf K}_{0} = \half \lpar
{\bf a}^{\dagger} {\bf a} - {\bf b}^{\dagger} {\bf b} \rpar$, with $\lbk {\bf K}_{-},{\bf K}_{+} \rbk = - 2 {\bf K}_{0}$
and $\lbk {\bf K}_{0},{\bf K}_{\pm} \rbk = \pm {\bf K}_{\pm}$; \\
(iii) the commutation relation between the operators ${\bf S}$ and ${\bf D}$ is null, i.e., $[ {\bf S},{\bf D} ] = 0$; and
consequently, \\
(iv) the Hamiltonian ${\bf H}$ simplifies to
\br
\lb{e4}
{\bf H}_{0} &=& \hbar \om \, {\bf S} + \half \hbar \om \, \sig_{{\rm z}} \; , \\
\lb{e5}
{\bf V} &=& \half \hbar \delta \, \sig_{{\rm z}} + \hbar \kappa_{{\rm eff}} \lpar {\bf A}^{\dagger} \sig_{-} + {\bf A}
\sig_{+} \rpar \; ,
\er
with $\lbk {\bf H}_{0},{\bf V} \rbk = 0$. This fact leads us to obtain the Hamiltonian ${\bf H}_{{\rm int}} = {\bf V}$ in
the interaction picture, which describes the well-known nonresonant JCM Hamiltonian for an atom interacting with the
quasi-mode ${\bf A}$, and whose coupling constant is given by $\kappa_{{\rm eff}}$. Thus, the unitary time-evolution
operator is the usual nonresonant JCM time-evolution operator.

If one considers the time-evolution operator $\bu(t) = \exp \lpar - \im {\bf V} t / \hbar \rpar$ of the atom-cavity system
written in the atomic basis, the elements $\bu_{{\rm ij}}(t)$ of the $2 \times 2$ matrix can be expressed as \cite{ref30}
\br
\lb{e6}
\bu_{11}(t) &=& \cos \lpar t \sqrt{\bet_{{\rm A}}} \; \rpar - \im \frac{\delta}{2} \frac{\sin \lpar t \sqrt{\bet_{{\rm A}}}
\; \rpar}{\sqrt{\bet_{{\rm A}}}} \; , \\
\lb{e7}
\bu_{12}(t) &=& - \im \kappa_{{\rm eff}} \; \frac{\sin \lpar t \sqrt{\bet_{{\rm A}}} \; \rpar}{\sqrt{\bet_{{\rm A}}}} \;
{\bf A} \; , \\
\lb{e8}
\bu_{21}(t) &=& - \im \kappa_{{\rm eff}} \; {\bf A}^{\dagger} \; \frac{\sin \lpar t \sqrt{\bet_{{\rm A}}} \; \rpar}
{\sqrt{\bet_{{\rm A}}}} \; , \\
\lb{e9}
\bu_{22}(t) &=& \cos \lpar t \sqrt{\vfi_{{\rm A}}} \; \rpar + \im \frac{\delta}{2} \frac{\sin \lpar t \sqrt{\vfi_{{\rm A}}}
\; \rpar}{\sqrt{\vfi_{{\rm A}}}} \; ,
\er
where $\vfi_{{\rm A}} = \kappa_{{\rm eff}}^{2} {\bf N}_{{\rm A}} + \lpar \frac{\delta}{2} \rpar^{2} {\bf 1}$ and
$\bet_{{\rm A}} = \vfi_{{\rm A}} + \kappa_{{\rm eff}}^{2} {\bf 1}$. This result permits us to determine the density operator
$\ro(t) = \bu(t) \ro(0) \bu^{\dagger}(t)$, being $\ro(0)$ the density operator of the system at time $t=0$. For convenience
in the calculations, we assume the atom is initially in the excited state and the cavity and external fields are in the
diagonal representation of coherent states, i.e., $\ro(0) = \ro_{{\rm at}}(0) \otimes \ro_{{\rm ab}}(0)$ with
$\ro_{{\rm at}}(0) = | e \rg \lg e |$ and
\be
\lb{e10}
\ro_{{\rm ab}}(0) = \ro_{{\rm a}}(0) \otimes \ro_{{\rm b}}(0) = \int \!\!\!\!\! \int \frac{d^{2} \alf_{{\rm a}} d^{2}
\alf_{{\rm b}}}{\pi^{2}} \, P_{{\rm a}}(\alf_{{\rm a}}) P_{{\rm b}}(\alf_{{\rm b}}) | \alf_{{\rm a}},\alf_{{\rm b}} \rg \lg
\alf_{{\rm a}},\alf_{{\rm b}} | \; ,
\ee
where $P(\alf)$ represents the Glauber-Sudarshan quasiprobability distribution for each field, and $| \alf_{{\rm a}},
\alf_{{\rm b}} \rg \equiv | \alf_{{\rm a}} \rg \otimes | \alf_{{\rm b}} \rg$. Consequently, the matrix elements
$\ro_{{\rm ij}}(t)$ can be calculated through the expressions
\br
\lb{e11}
\ro_{11}(t) &=& \bu_{11}(t) \; \ro_{{\rm ab}}(0) \; \bu_{11}^{\dagger}(t) \; , \\
\lb{e12}
\ro_{12}(t) &=& \bu_{11}(t) \; \ro_{{\rm ab}}(0) \; \bu_{21}^{\dagger}(t) \; , \\
\lb{e13}
\ro_{21}(t) &=& \bu_{21}(t) \; \ro_{{\rm ab}}(0) \; \bu_{11}^{\dagger}(t) \; , \\
\lb{e14}
\ro_{22}(t) &=& \bu_{21}(t) \; \ro_{{\rm ab}}(0) \; \bu_{21}^{\dagger}(t) \; .
\er
Note that $\ro(t)$ describes the exact solution of the Schr\"{o}dinger equation in the interaction picture with the
nonresonant driven-JCM Hamiltonian. Using this solution we can establish analytical expressions for the time evolution of
various functions characterizing the quantum state of the cavity field, such as the atomic inversion, the moments
associated to the photon-number operator, the Mandel's $Q$ parameter, the photon-number distribution and its respective
entropy, the variances of the quadrature components, and the Wigner function. For simplicity, the initial state of the
driving field will be fix in the coherent state throughout this paper $\lpar \ro_{{\rm b}}(0) = | \beta \rg \lg \beta |
\rpar$, and the initial state of the cavity field will assume two different possibilities: the even- and odd-coherent
states \cite{ref29}. With respect to Glauber-Sudarshan quasiprobability distribution, these considerations are equivalent
to $P_{{\rm b}}^{({\rm c})}(\alf_{{\rm b}}) = \pi \delta^{(2)}(\alf_{{\rm b}} - \beta)$ and
\br
P_{{\rm a}}^{({\rm e})}(\alf_{{\rm a}}) &=& \frac{\pi}{4} \frac{\exp \lpar | \alf_{{\rm a}} |^{2} \rpar}{\cosh \lpar
| \alf |^{2} \rpar} \lbk \delta^{(2)} (\alf_{{\rm a}} - \alf) + \delta^{(2)} (\alf_{{\rm a}} + \alf) + 2 \cosh \lpar \alf
\frac{\partial}{\partial \alf_{{\rm a}}} - \alf^{\ast} \frac{\partial}{\partial \alf_{{\rm a}}^{\ast}} \rpar \delta^{(2)}
(\alf_{{\rm a}}) \rbk \; , \nn \\
P_{{\rm a}}^{({\rm o})}(\alf_{{\rm a}}) &=& \frac{\pi}{4} \frac{\exp \lpar | \alf_{{\rm a}} |^{2} \rpar}{\sinh \lpar
| \alf |^{2} \rpar} \lbk \delta^{(2)} (\alf_{{\rm a}} - \alf) + \delta^{(2)} (\alf_{{\rm a}} + \alf) - 2 \cosh \lpar \alf
\frac{\partial}{\partial \alf_{{\rm a}}} - \alf^{\ast} \frac{\partial}{\partial \alf_{{\rm a}}^{\ast}} \rpar \delta^{(2)}
(\alf_{{\rm a}}) \rbk \; , \nn
\er
being $\delta^{(2)}(z)$ the two-dimensional delta function. According to Glauber \cite{ref28}: `` If the singularities of
$P(\alf)$ are of types stronger than those of delta function, e.g., derivatives of delta function, the field represented
will have no classical analog". Thus, in the next sections we will investigate the influence of the amplitude of the
driving field and detuning parameters on the nonclassical effects of the cavity field where, in particular, the atomic
inversion and the Wigner function should be emphasized.

\section{Atomic inversion}

The atomic inversion ${\cal I}(t) \equiv \tr \lbk \ro(t) \sig_{{\rm z}} \rbk$ is a quantity of central interest in this
section since it is easily accessible in experiments \cite{ref31}. For the atom-cavity system described in the previous
section, this function can be written in an integral form as (see appendix A for calculational details)
\be
\lb{e15}
{\cal I}(t) = \int \!\!\!\!\! \int \frac{d^{2} \alf_{{\rm a}} d^{2} \alf_{{\rm b}}}{\pi^{2}} \, P_{{\rm a}}(\alf_{{\rm a}})
P_{{\rm b}}(\alf_{{\rm b}}) \, \Xi ( \alf_{{\rm a}},\alf_{{\rm b}};t ) \; ,
\ee
where
\bd
\Xi (\alf_{{\rm a}},\alf_{{\rm b}};t) = 1 - 2 \exp \lpar - \left| \eps_{{\rm a}} \alf_{{\rm a}} + \eps_{{\rm b}}
\alf_{{\rm b}} \right|^{2} \rpar \sum_{n=0}^{\infty} \frac{\left| \eps_{{\rm a}} \alf_{{\rm a}} + \eps_{{\rm b}}
\alf_{{\rm b}} \right|^{2n}}{n!} |G_{n}(t)|^{2} \; .
\ed
Here, the function $G_{n}(t) = - \im (\Omega_{n} / \Delta_{n}) \sin (\Delta_{n} t/2)$ is responsible for the time
evolution of the atomic inversion, being $\Delta_{n}^{2} = \delta^{2} + \Omega_{n}^{2}$ and $\Omega_{n} = 2
\kappa_{{\rm eff}} \sqrt{n+1}$ the effective Rabi frequency. Note that the Eq. (\ref{e15}) can be obtained for any states
of the cavity and external electromagnetic fields. For instance, if one considers the both cavity and external fields in
the coherent states, the atomic inversion coincides with $\Xi (\alf,\beta;t)$. This situation was investigated by Dutra
et al. \cite{ref17} for the atomic excitation probability $P_{{\rm e}}(t) = \half [{\cal I}(t) + 1]$ and $\delta = 0$
(resonance condition), where the authors have shown that $P_{{\rm e}}(t)$ is connected to Wigner characteristic function
of the cavity field since the conditions $\kappa_{{\rm a}} \gg \kappa_{{\rm b}}$, $\kappa_{{\rm b}} t \ll 1$, and
$| \beta | \gg (\kappa_{{\rm a}} / \kappa_{{\rm b}}) \kappa_{{\rm a}} t$ (intense driving field) are satisfied. On the
other hand, if one considers the cavity and external fields in the thermal and coherent states, respectively, the atomic
inversion is given by
\be
\lb{e16}
{\cal I}_{{\rm th}}(t) = 1 - \frac{2}{1 + \eps_{{\rm a}}^{2} \bar{n}} \exp \lpar - \frac{\eps_{{\rm b}}^{2} | \beta |^{2}}
{1 + \eps_{{\rm a}}^{2} \bar{n}} \rpar \sum_{n=0}^{\infty} \lpar \frac{\eps_{{\rm a}}^{2} \bar{n}}{1 + \eps_{{\rm a}}^{2}
\bar{n}} \rpar^{n} L_{n} \lbk - \frac{\eps_{{\rm b}}^{2} | \beta |^{2}}{\eps_{{\rm a}}^{2} \bar{n} (1 + \eps_{{\rm a}}^{2}
\bar{n})} \rbk |G_{n}(t)|^{2} \; .
\ee
In this expression, $\bar{n}$ is the mean number of thermal photons at time $t=0$, and $L_{n}(z)$ corresponds to a Laguerre
polynomial. Furthermore, the parameter $\eps_{{\rm a}({\rm b})}$ represents a scale factor for $\bar{n}$ $(| \beta |)$. It
is important mentioning that Eq. (\ref{e16}) corroborates the numerical investigations realized by Li and Gao \cite{ref21}
for the thermal states, and this fact leads us to proceed with the study of atomic inversion for the even- and odd-coherent 
states.
\begin{figure}[!t]
\centering
\begin{minipage}[b]{0.45\linewidth}
\includegraphics[width=\linewidth]{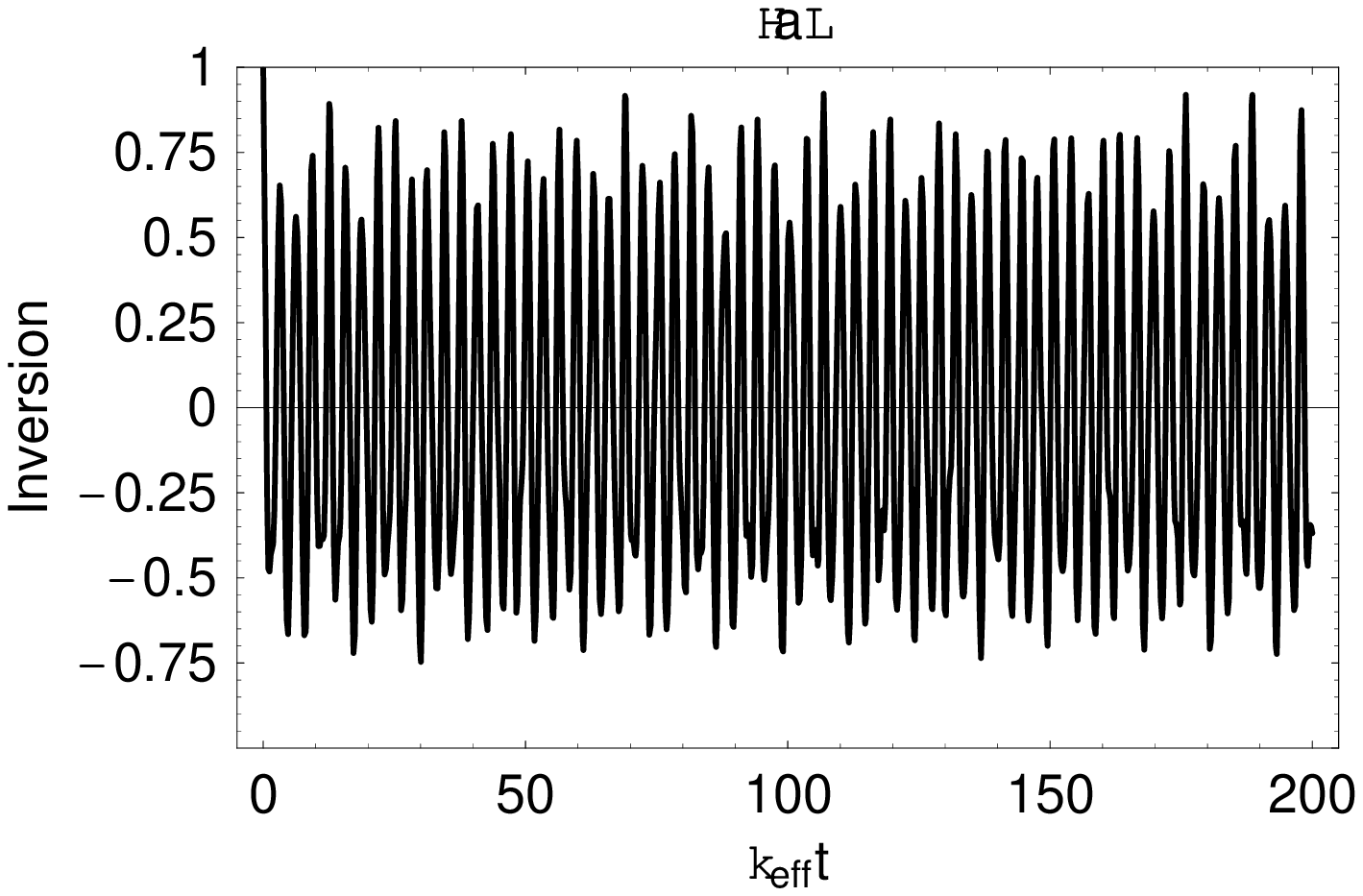}
\end{minipage} \hfill
\begin{minipage}[b]{0.45\linewidth}
\includegraphics[width=\linewidth]{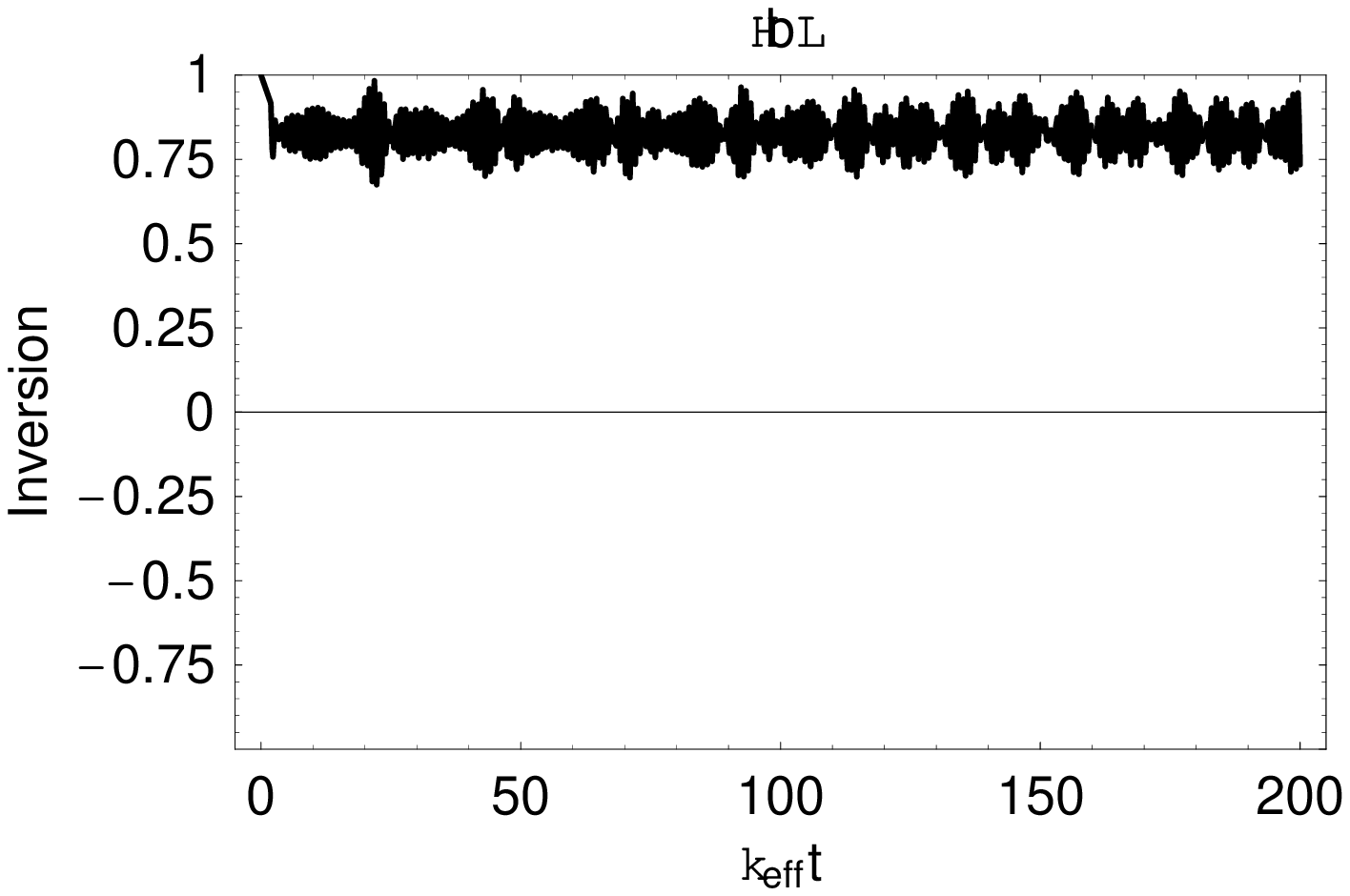}
\end{minipage} \hfill
\begin{minipage}[b]{0.45\linewidth}
\includegraphics[width=\linewidth]{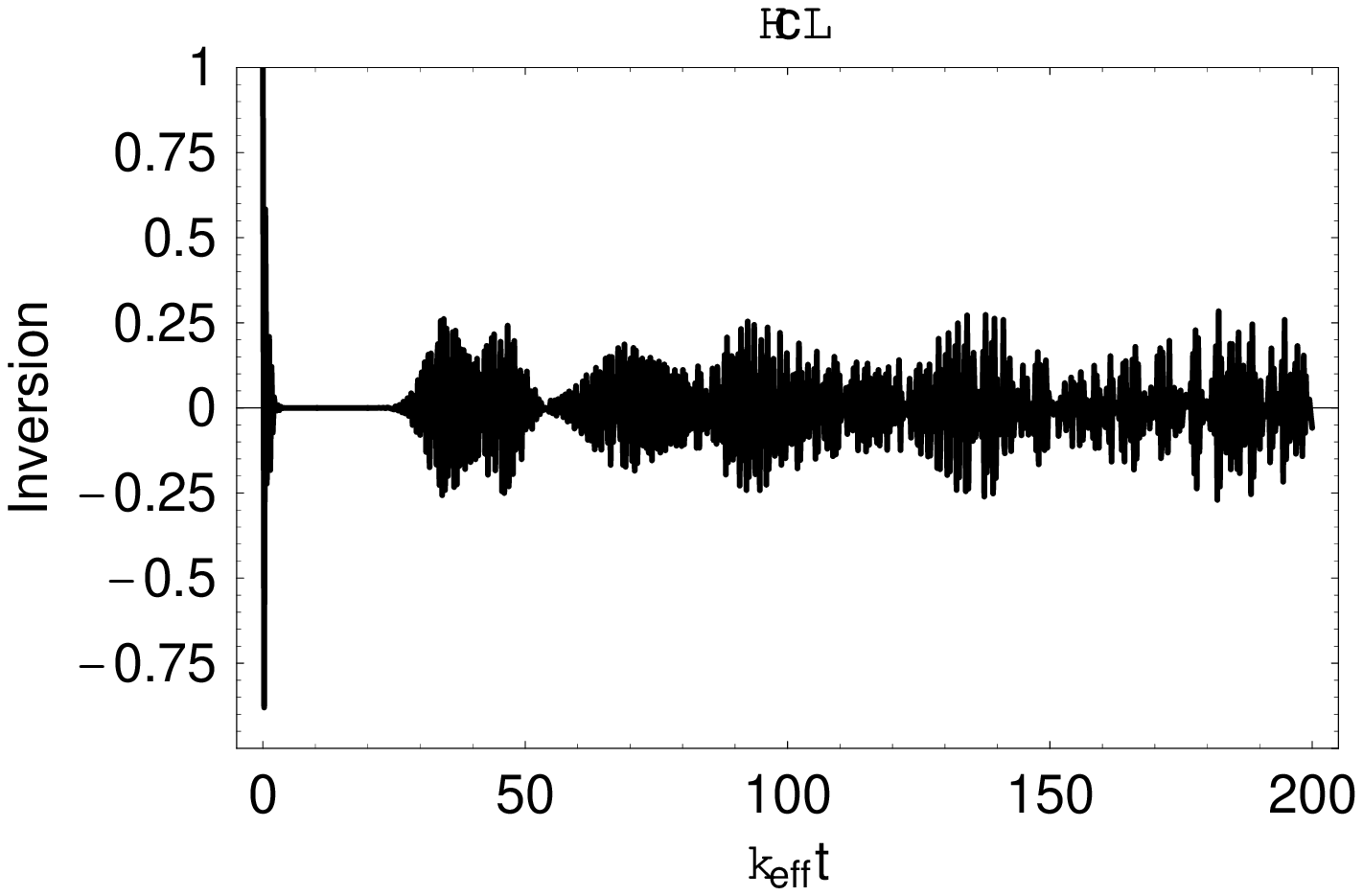}
\end{minipage} \hfill
\begin{minipage}[b]{0.45\linewidth}
\includegraphics[width=\linewidth]{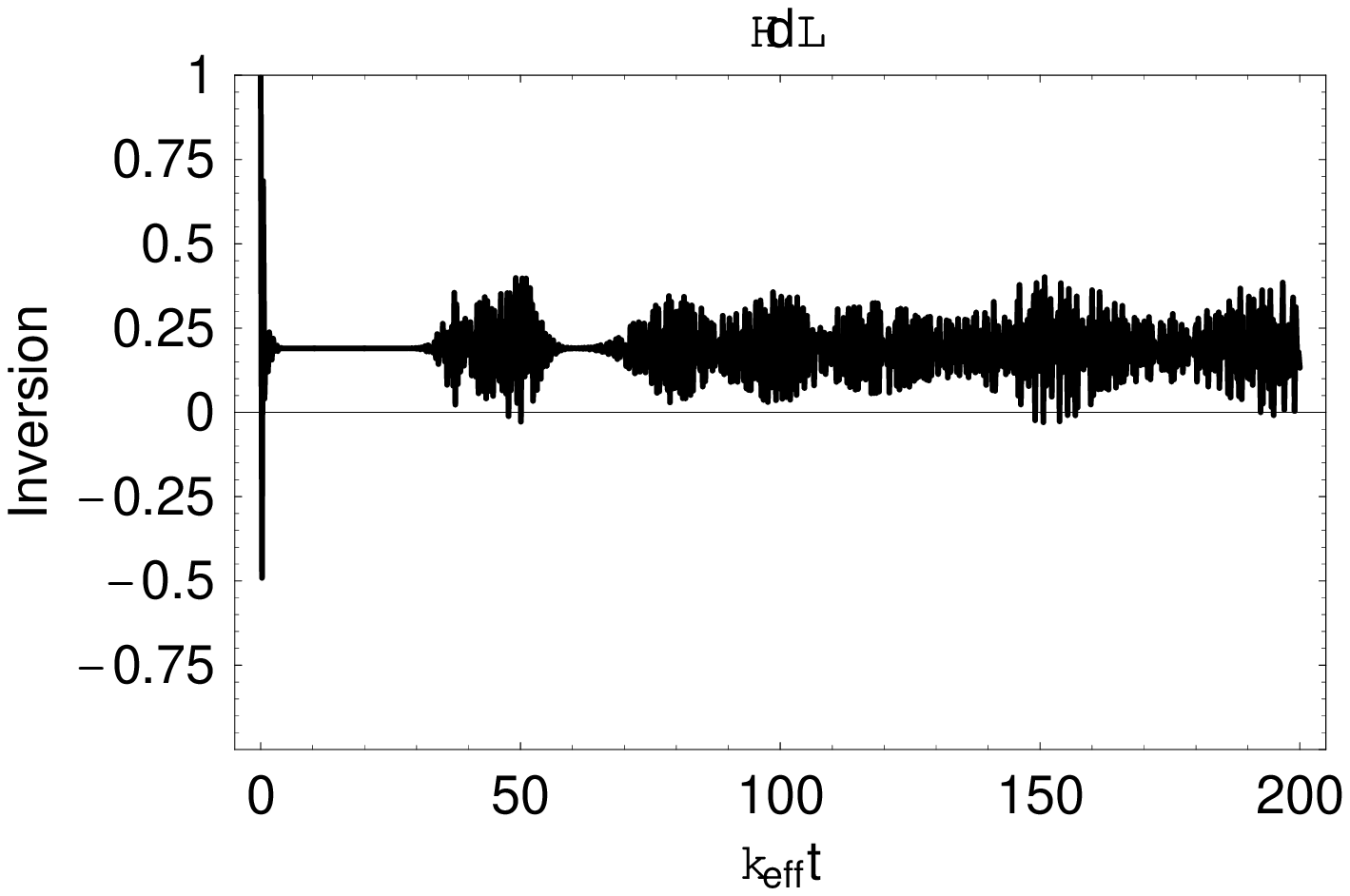}
\end{minipage}
\caption{Time evolution of the atomic inversion ${\cal I}_{{\rm e}}(t)$ of the atom initially prepared in the excited
state interacting with cavity and external fields in the even-coherent and coherent states, respectively. These pictures
correspond to (a,c) $\delta = 0$ (resonant) and (b,d) $\delta = 6 \kappa_{{\rm eff}}$ (nonresonant) for $\eps_{{\rm a}} =
3/ \sqrt{10}$, $\eps_{{\rm b}} = 1/ \sqrt{10}$, and $| \alf | = 1$ fixed, where two different values of amplitude of the
driving field were considered: (a,b) $| \beta | = 2$ and (c,d) $| \beta | = 20$.}
\end{figure}
\begin{figure}[!t]
\centering
\begin{minipage}[b]{0.45\linewidth}
\includegraphics[width=\linewidth]{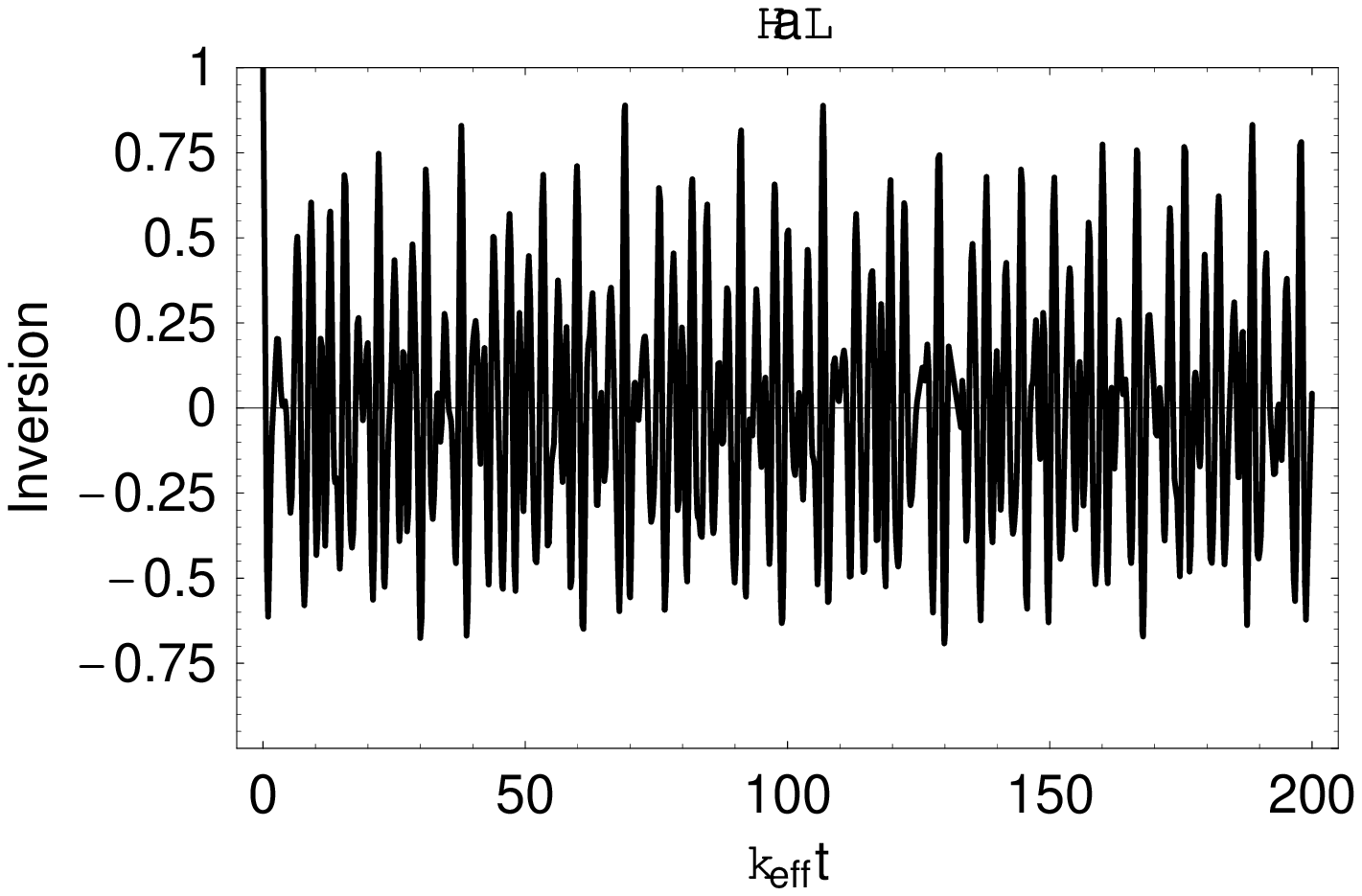}
\end{minipage} \hfill
\begin{minipage}[b]{0.45\linewidth}
\includegraphics[width=\linewidth]{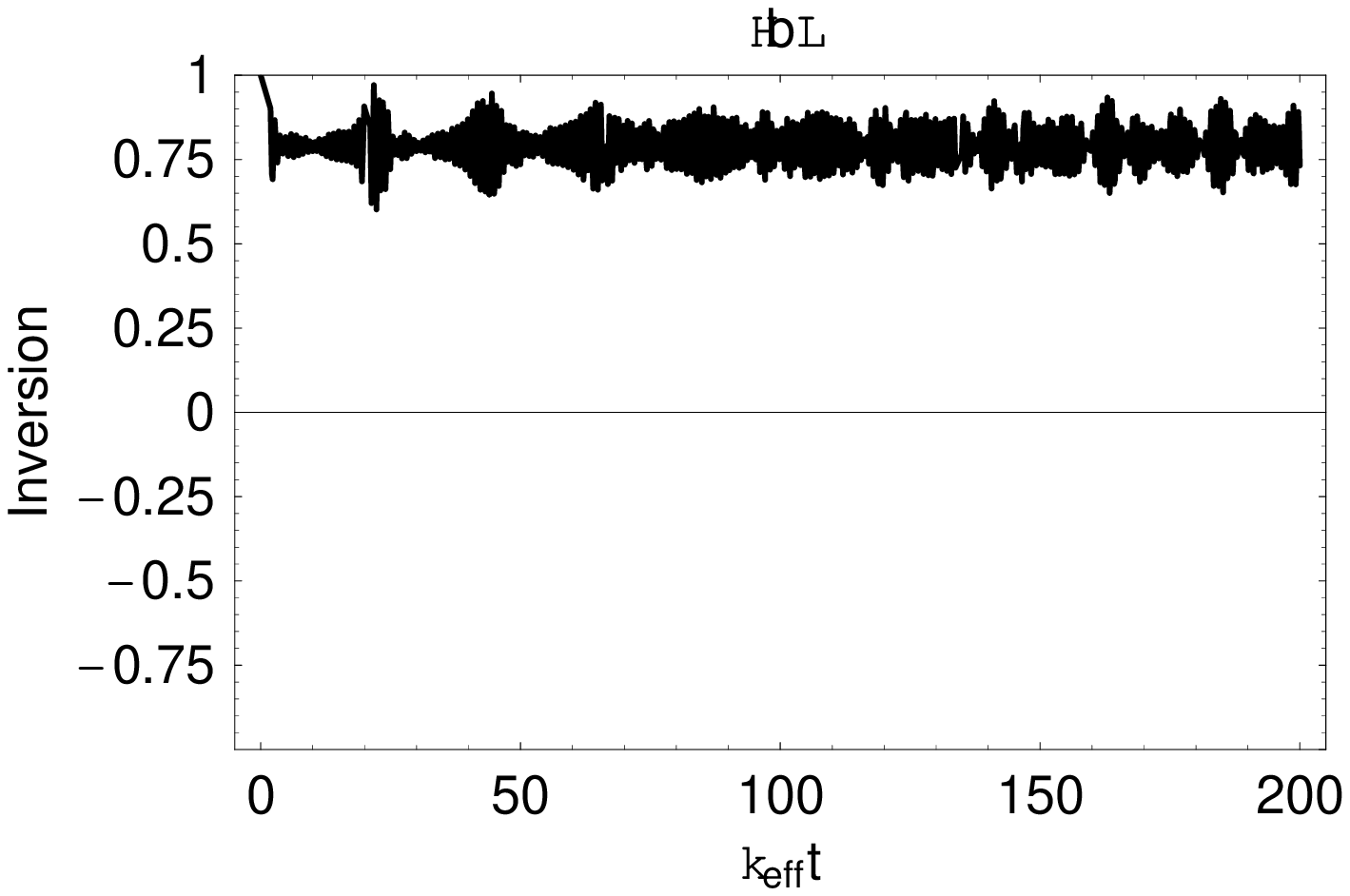}
\end{minipage} \hfill
\begin{minipage}[b]{0.45\linewidth}
\includegraphics[width=\linewidth]{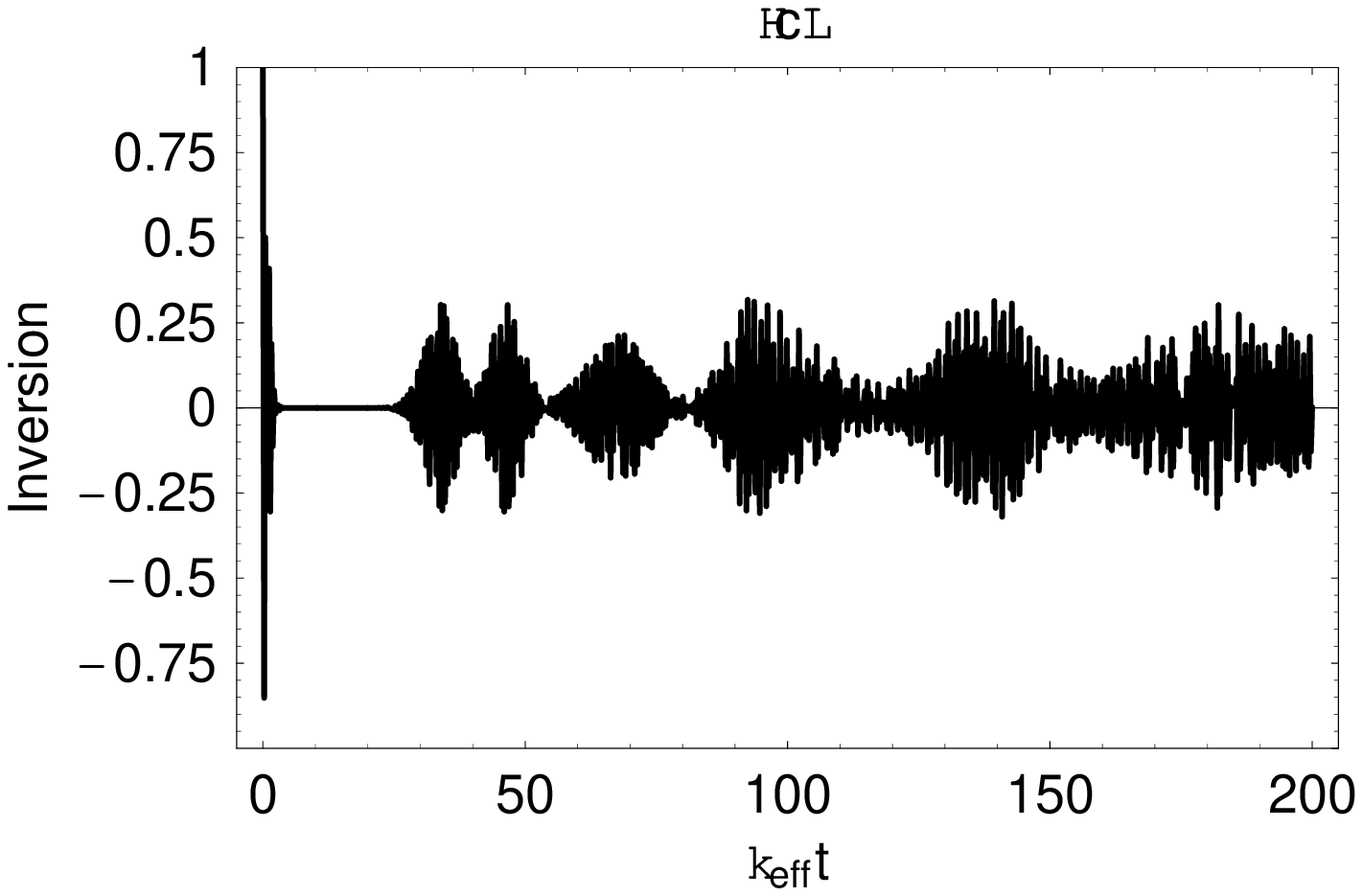}
\end{minipage} \hfill
\begin{minipage}[b]{0.45\linewidth}
\includegraphics[width=\linewidth]{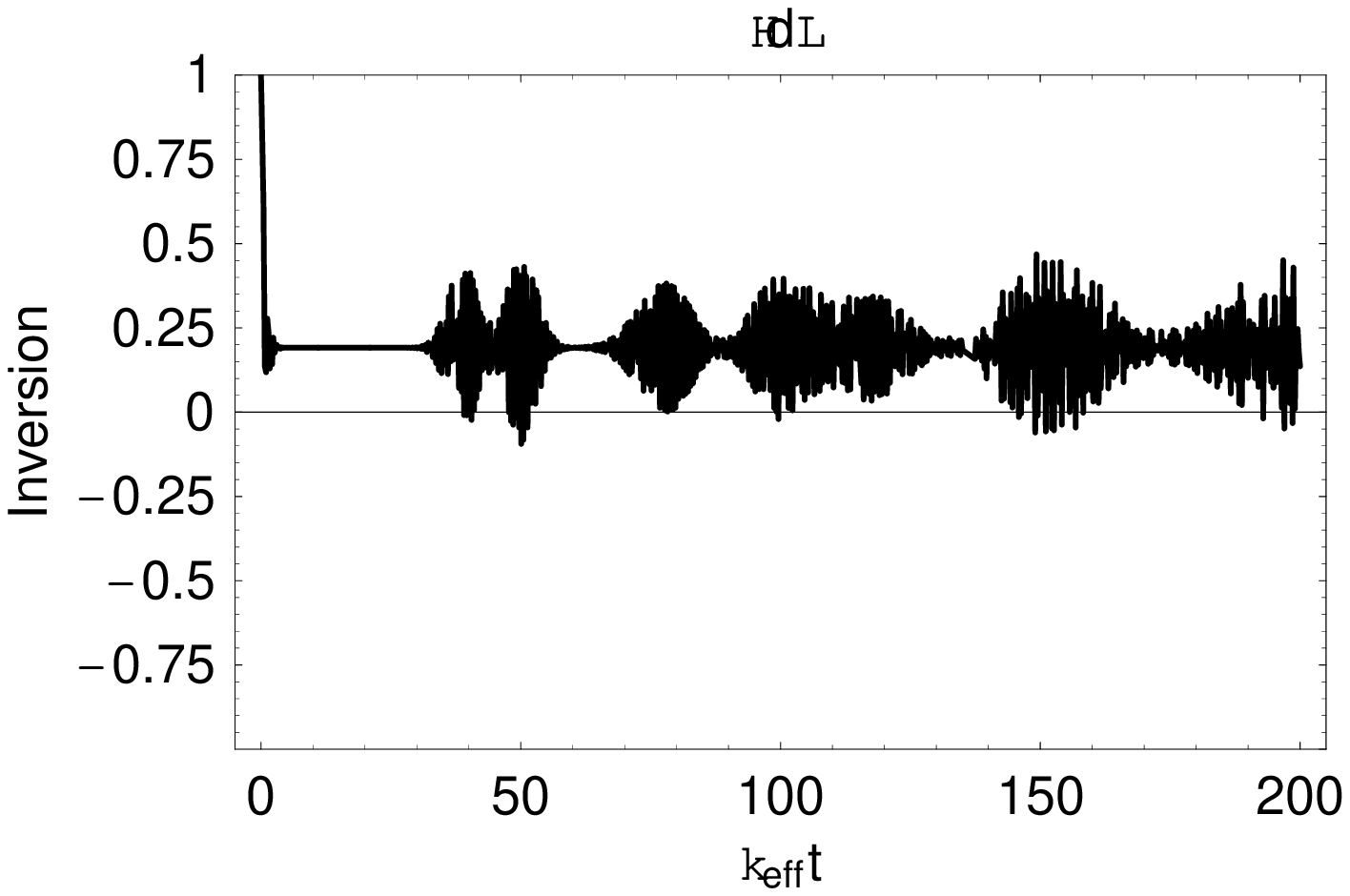}
\end{minipage}
\caption{Plots of ${\cal I}_{{\rm o}}(t)$ versus $\kappa_{{\rm eff}} t \in [0,200]$ for (a,c) $\delta = 0$ (resonant) and
(b,d) $\delta = 6 \kappa_{{\rm eff}}$ (nonresonant), with $\eps_{{\rm a}} = 3/ \sqrt{10}$, $\eps_{{\rm b}}=1/ \sqrt{10}$,
and $| \alf | = 1$ fixed. In both situations were considered different values of amplitude of the driving field, i.e.,
(a,b) $| \beta | = 2$ and (c,d) $| \beta | = 20$.}
\end{figure}

Let us consider the Glauber-Sudarshan quasiprobability distributions $P_{{\rm a}}^{({\rm e})}(\alf_{{\rm a}})$ and
$P_{{\rm a}}^{({\rm o})}(\alf_{{\rm a}})$ for the even- and odd-coherent states into the Eq. (\ref{e15}), whose integrals 
in the complex $\alf_{{\rm a}}$- and $\alf_{{\rm b}}$-planes can be evaluated without technical difficulties. In both
situations, the atomic inversion is expressed in the compact form
\br
\lb{e17}
{\cal I}_{{\rm e}}(t) &=& 1 - 2 \sum_{n=0}^{\infty} {\mathfrak F}_{n}^{({\rm e})}(\alf,\beta) |G_{n}(t)|^{2} \; , \\
\lb{e18}
{\cal I}_{{\rm o}}(t) &=& 1 - 2 \sum_{n=0}^{\infty} {\mathfrak F}_{n}^{({\rm o})}(\alf,\beta) |G_{n}(t)|^{2} \; ,
\er
with ${\mathfrak F}_{n}^{({\rm e})}(\alf,\beta)$ and ${\mathfrak F}_{n}^{({\rm o})}(\alf,\beta)$ given by
\br
{\mathfrak F}_{n}^{({\rm e})}(\alf,\beta) &=& \frac{\exp \lpar | \alf |^{2} \rpar}{4 \cosh \lpar | \alf |^{2} \rpar}
\lbr \exp \lpar - \left| \eps_{{\rm a}} \alf + \eps_{{\rm b}} \beta \right|^{2} \rpar \frac{\left| \eps_{{\rm a}} \alf +
\eps_{{\rm b}} \beta \right|^{2n}}{n!} + \exp \lpar - \left| \eps_{{\rm a}} \alf - \eps_{{\rm b}} \beta \right|^{2} \rpar
\frac{\left| \eps_{{\rm a}} \alf - \eps_{{\rm b}} \beta \right|^{2n}}{n!} \right. \nn \\
& & \left. + \, 2 \exp \lpar - 2 | \alf |^{2} \rpar \re \lbk \exp \lbk \lpar \eps_{{\rm a}} \alf + \eps_{{\rm b}} \beta
\rpar \lpar \eps_{{\rm a}} \alf - \eps_{{\rm b}} \beta \rpar^{\ast} \rbk \frac{\lbk - \lpar \eps_{{\rm a}} \alf +
\eps_{{\rm b}} \beta \rpar \lpar \eps_{{\rm a}} \alf - \eps_{{\rm b}} \beta \rpar^{\ast} \rbk^{n}}{n!} \rbk \rbr \; , \nn \\
{\mathfrak F}_{n}^{({\rm o})}(\alf,\beta) &=& \frac{\exp \lpar | \alf |^{2} \rpar}{4 \sinh \lpar | \alf |^{2} \rpar}
\lbr \exp \lpar - \left| \eps_{{\rm a}} \alf + \eps_{{\rm b}} \beta \right|^{2} \rpar \frac{\left| \eps_{{\rm a}} \alf +
\eps_{{\rm b}} \beta \right|^{2n}}{n!} + \exp \lpar - \left| \eps_{{\rm a}} \alf - \eps_{{\rm b}} \beta \right|^{2} \rpar
\frac{\left| \eps_{{\rm a}} \alf - \eps_{{\rm b}} \beta \right|^{2n}}{n!} \right. \nn \\
& & \left. - \, 2 \exp \lpar - 2 | \alf |^{2} \rpar \re \lbk \exp \lbk \lpar \eps_{{\rm a}} \alf + \eps_{{\rm b}} \beta
\rpar \lpar \eps_{{\rm a}} \alf - \eps_{{\rm b}} \beta \rpar^{\ast} \rbk \frac{\lbk - \lpar \eps_{{\rm a}} \alf +
\eps_{{\rm b}} \beta \rpar \lpar \eps_{{\rm a}} \alf - \eps_{{\rm b}} \beta \rpar^{\ast} \rbk^{n}}{n!} \rbk \rbr \; . \nn
\er
Fig. 2 shows the plots of ${\cal I}_{{\rm e}}(t)$ versus $\kappa_{{\rm eff}} t$ when the atom-cavity system is resonant
(a,c) $\delta = 0$ and nonresonant (b,d) $\delta = 6 \kappa_{{\rm eff}}$ for $\eps_{{\rm a}} = 3/ \sqrt{10}$,
$\eps_{{\rm b}} = 1/ \sqrt{10}$, and $| \alf | = 1$ fixed, with two different values of amplitude of the driving field:
(a,b) $| \beta | = 2$ and (c,d) $| \beta | = 20$. Since the atom was initially prepared in the excited state, the value of
the atomic inversion at the time origin is equal to one in all situations. In Fig. 2(a), we can perceive that
${\cal I}_{{\rm e}}(t)$ behaves in a fairly irregular manner and the revivals are not well defined (in particular, the
revivals are considered as a manifestation of the quantum nature of the electromagnetic field inside the cavity); while in
Fig. 2(c), the collapses and revivals appear when the driving field is strong. Now, if one analyses the Figs. 2(b) and (d)
we conclude that the collapses and revivals can be controlled by the detuning between the cavity (external) field and the
atomic transition (in particular, the revivals have a regular structure and small amplitude). Similarly, Fig. 3 shows the
plots of ${\cal I}_{{\rm o}}(t)$ versus $\kappa_{{\rm eff}} t$ considering the same parameter set used in the previous
figure, where we verify that: (i) different structures of collapses and revivals are present, and (ii) the effects of the
parameters $| \beta |$ and $\delta$ on the atomic inversion ${\cal I}_{{\rm o}}(t)$ are completely analogous to the
even-coherent states. G\'{o}ra and Jedrzejek \cite{ref32} have shown that in the usual JCM with the cavity field prepared
initially in a coherent state with a small mean number of photons (i.e., $\lg {\bf n} \rg_{{\rm c}} \approx 2$ at time
$t=0$), the atomic inversion displays distinct collapses and revivals provided the atom and the field are slightly detuned,
and the {\em long-time} behaviour of the model presents superstructures such as fractional revivals and superrevivals. In
this sense, the Figs. 2(b) and 3(b) present a {\em short-period} behaviour with analogous superstructures and this fact is
associated to the small mean number of photons used for both the cavity and external fields ($\lg {\bf n}_{{\rm a}} (0)
\rg_{{\rm e}} \approx 0.762$ and $\lg {\bf n}_{{\rm a}}(0) \rg_{{\rm o}} \approx 1.313$, with $\lg {\bf n}_{{\rm b}} (0)
\rg_{{\rm c}} \approx 4$ fixed), since the detuning is large as compared to the effective coupling constant (e.g., $\delta
/ 2 \kappa_{{\rm eff}} =3$). Moreover, these superstructures disappear when we consider $\lg {\bf n}_{{\rm b}} (0)
\rg_{{\rm c}} \approx 400$ in Figs. 2(d) and 3(d). Summarizing, the amplitude of the driving field and the detuning
parameter have a strong influence on the structures of collapses and revivals in the driven JCM, and this fact leads us to
investigate its effects on the nonclassical properties of the cavity field via Wigner function.

\section{Wigner function}

\begin{figure}[!t]
\centering
\begin{minipage}[b]{0.45\linewidth}
\includegraphics[width=\linewidth]{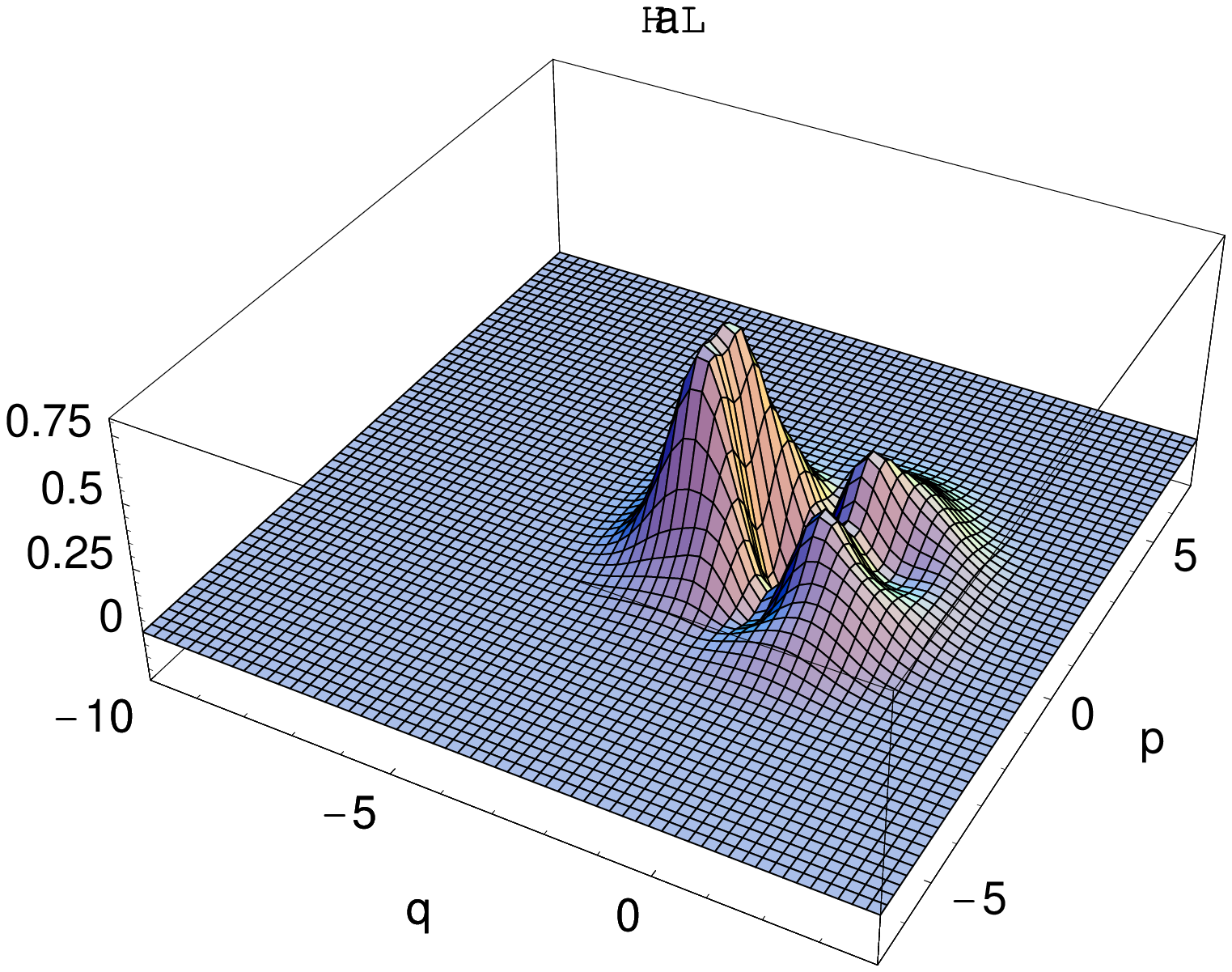}
\end{minipage} \hfill
\begin{minipage}[b]{0.45\linewidth}
\includegraphics[width=\linewidth]{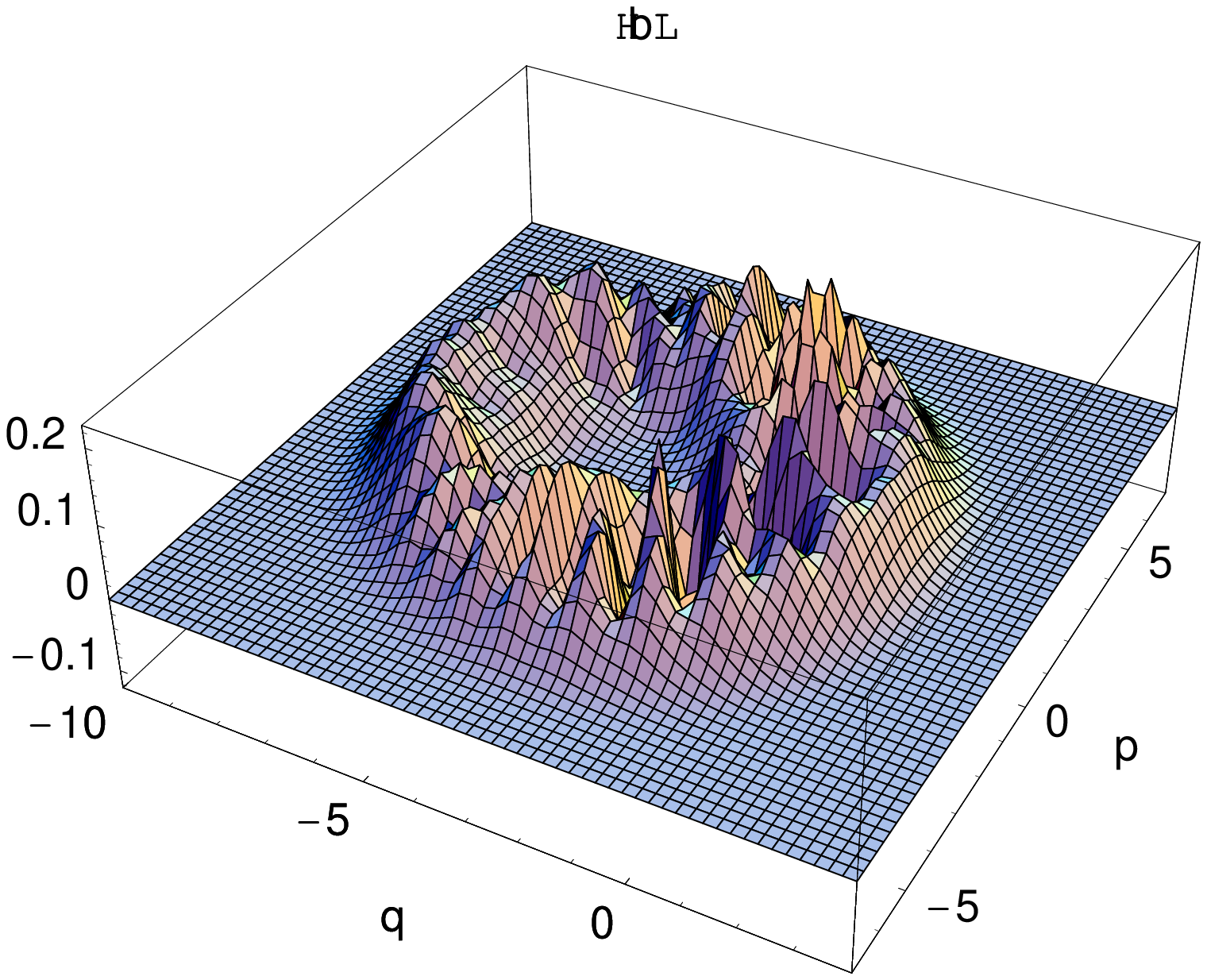}
\end{minipage} \hfill
\begin{minipage}[b]{0.45\linewidth}
\includegraphics[width=\linewidth]{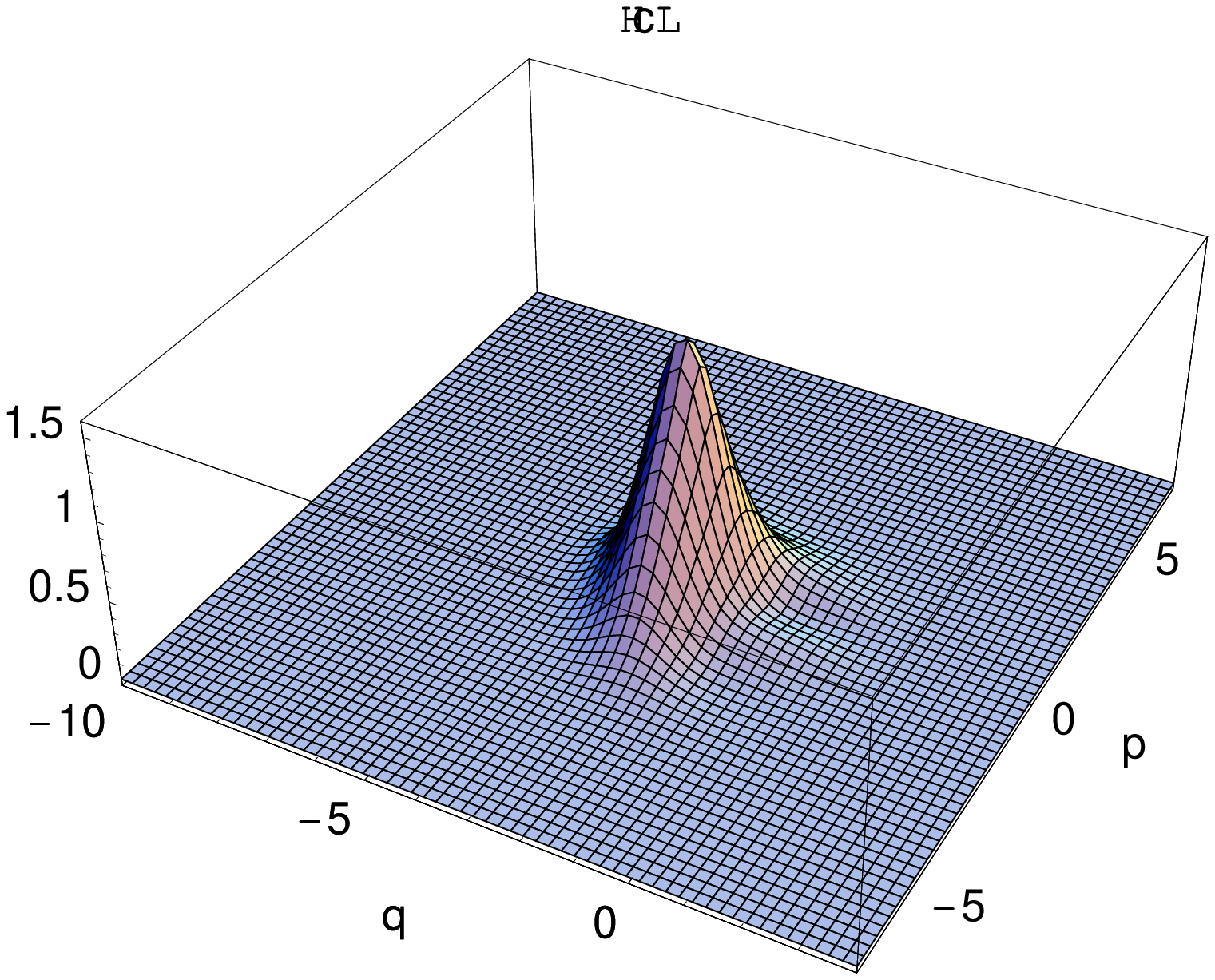}
\end{minipage} \hfill
\begin{minipage}[b]{0.45\linewidth}
\includegraphics[width=\linewidth]{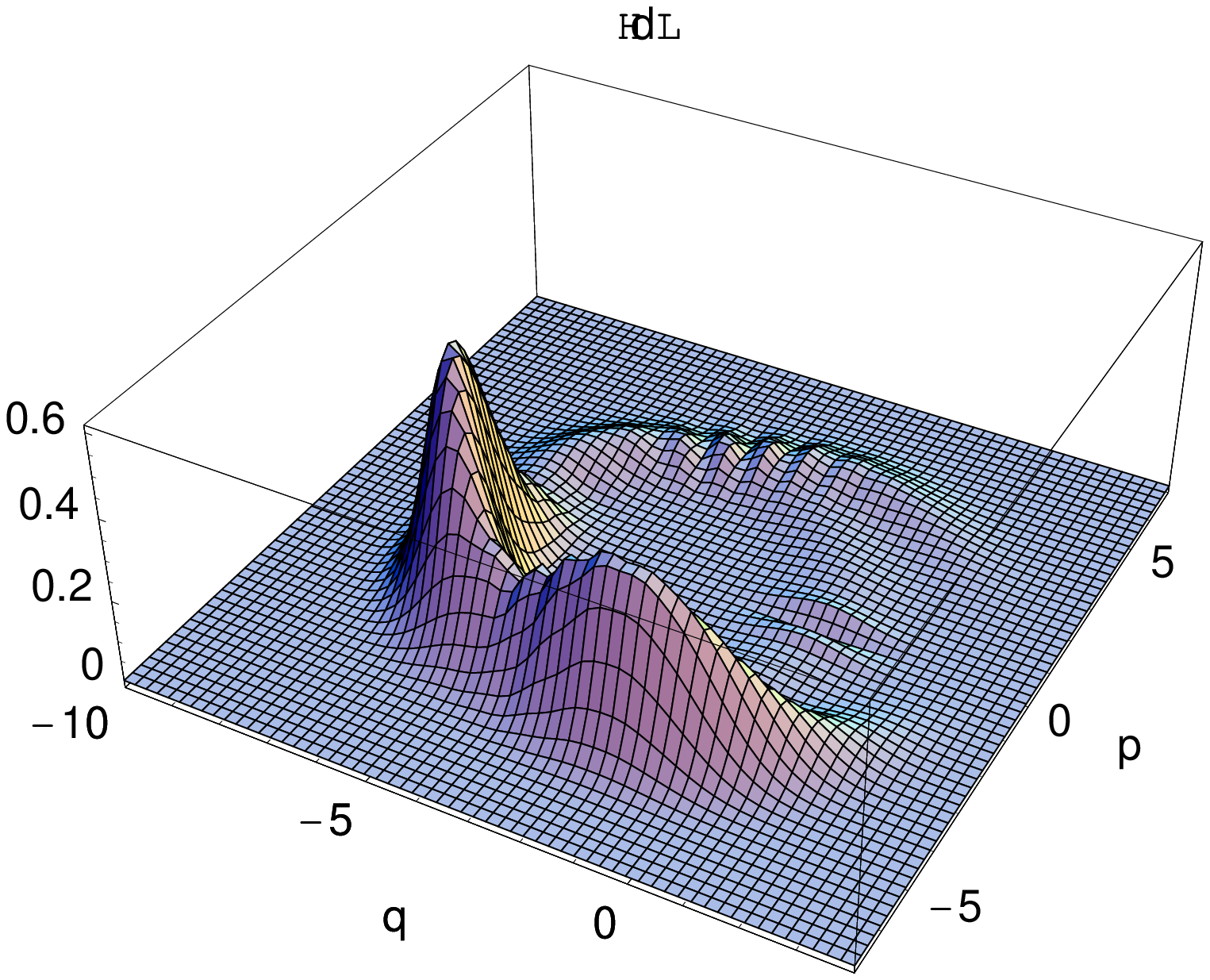}
\end{minipage}
\caption{Plots of $W_{{\rm a}}^{({\rm e})}(\gamma;t)$ versus $p \in [-7,7]$ and $q \in [-10,4]$ for the atom-cavity
system with two different values of detuning: (a,b) $\delta = 0$ (resonant) and (c,d) $\delta = 10 \kappa_{{\rm eff}}$
(nonresonant), where the parameters $| \alf | = 1$ $( \lg {\bf n}_{{\rm a}} \rg_{{\rm e}} \approx 0.762 )$ and
$\kappa_{{\rm eff}} t = 100$ were fixed in the present simulation. In both situations, the condition
$\kappa_{{\rm a}({\rm b})} = \kappa$ was established and the values of amplitude of the driving field (a,c) $| \beta | = 2$
$( \lg {\bf n}_{{\rm b}} \rg_{{\rm c}} = 4 )$ and (b,d) $| \beta | = 5$ $(\lg {\bf n}_{{\rm b}} \rg_{{\rm c}} = 25 )$
considered.}
\end{figure}
In many recent textbooks on quantum optics \cite{ref35}, the Wigner function is generally defined in terms of an auxiliary
function (also denominated as Wigner characteristic function) which describes the symmetric ordering of creation and
annihilation operators of the electromagnetic field, i.e., $\chi (\xi) \equiv \tr [ \ro {\bf D} (\xi) ]$ with ${\bf D}(\xi)
= \exp \lpar \xi {\bf a}^{\dagger} - \xi^{\ast} {\bf a} \rpar$ being the displacement operator. The connection between both
functions is established by means of a two-dimensional Fourier transform as follows:
\be
\lb{e19}
W(\gamma) = \int \frac{d^{2} \xi}{\pi} \exp \lpar \gamma \xi^{\ast} - \gamma^{\ast} \xi \rpar \chi (\xi) \; .
\ee
Thus, if one considers the cavity field in the framework of the driven JCM, its Wigner characteristic function can be
defined in a similar form to atomic inversion,
\be
\lb{e20}
\chi (\xi;t) = \int \!\!\!\!\! \int \frac{d^{2} \alf_{{\rm a}} d^{2} \alf_{{\rm b}}}{\pi^{2}} \, P_{{\rm a}}(\alf_{{\rm a}})
P_{{\rm b}}(\alf_{{\rm b}}) \, \widetilde{{\rm K}}_{\xi} ( \alf_{{\rm a}},\alf_{{\rm b}};t ) \; ,
\ee
with $\widetilde{{\rm K}}_{\xi} ( \alf_{{\rm a}},\alf_{{\rm b}};t )$ given by
\bd
\widetilde{{\rm K}}_{\xi} ( \alf_{{\rm a}},\alf_{{\rm b}};t ) = \lg \alf_{{\rm a}},\alf_{{\rm b}} | \bu_{11}^{\dagger}(t)
{\bf D}_{{\rm a}}(\xi) \bu_{11}(t) | \alf_{{\rm a}},\alf_{{\rm b}} \rg + \lg \alf_{{\rm a}},\alf_{{\rm b}} |
\bu_{21}^{\dagger}(t) {\bf D}_{{\rm a}}(\xi) \bu_{21}(t) | \alf_{{\rm a}},\alf_{{\rm b}} \rg \; .
\ed
Here, the displacement operator ${\bf D}_{{\rm a}}(\xi)$ is associated with the cavity field. Now, substituting
$\chi (\xi;t)$ into Eq. (\ref{e19}), the expression for the Wigner function is promptly obtained,
\be
\lb{e21}
W_{{\rm a}}(\gamma;t) = \int \!\!\!\!\! \int \frac{d^{2} \alf_{{\rm a}} d^{2} \alf_{{\rm b}}}{\pi^{2}} \, P_{{\rm a}}
(\alf_{{\rm a}}) P_{{\rm b}}(\alf_{{\rm b}}) \, {\rm K}_{\gamma} ( \alf_{{\rm a}},\alf_{{\rm b}};t ) \; ,
\ee
where the label $\gamma$ corresponds to representation in the complex phase-space and
\be
\lb{e22}
{\rm K}_{\gamma}(\alf_{{\rm a}},\alf_{{\rm b}};t) = \int \frac{d^{2} \xi}{\pi} \exp \lpar \gamma \xi^{\ast} - \gamma^{\ast}
\xi \rpar \widetilde{{\rm K}}_{\xi}(\alf_{{\rm a}},\alf_{{\rm b}};t) \; .
\ee
The functions $\widetilde{{\rm K}}_{\xi}(\alf_{{\rm a}},\alf_{{\rm b}};t)$ and ${\rm K}_{\gamma}(\alf_{{\rm a}},
\alf_{{\rm b}};t)$ were derived with details in the appendix B. In particular, when $t=0$ the function ${\rm K}_{\gamma}
(\alf_{{\rm a}},\alf_{{\rm b}};0) = 2 \exp ( - 2 | \gamma - \alf_{{\rm a}} |^{2})$ does not depend of variables associated
with the external field and this fact leads us to write the initial Wigner function as
\bd
W_{{\rm a}}(\gamma;0) = 2 \int \frac{d^{2} \alf_{{\rm a}}}{\pi} \exp ( - 2 | \gamma - \alf_{{\rm a}} |^{2}) P_{{\rm a}}
(\alf_{{\rm a}}) \; .
\ed
This expression represents a Gaussian smoothing process of the integrand $P_{{\rm a}}(\alf_{{\rm a}})$ such that
$W_{{\rm a}}(\gamma;0)$ is a well-defined function in the phase space $p = \sqrt{2} \, \ima (\gamma)$ and $q = \sqrt{2} \,
\re (\gamma)$. On the other hand, for $t > 0$ the function ${\rm K}_{\gamma}(\alf_{{\rm a}},\alf_{{\rm b}};t)$ is
responsible for the entanglement between the cavity and external fields (here represented by the Glauber-Sudarshan
quasiprobability distributions $P_{{\rm a}}(\alf_{{\rm a}})$ and $P_{{\rm b}}(\alf_{{\rm b}})$, respectively) since the
complex variables $\alf_{{\rm a}}$ and $\alf_{{\rm b}}$ are completely correlated. Furthermore, it is important mentioning
that $\chi (\xi;t)$ and $W_{{\rm a}}(\gamma;t)$ can be evaluated for any states of the cavity and external fields (similar
condition was established for atomic inversion) without restrictions on the different interaction times, and the expressions
obtained analytically from this procedure generalize the results previously discussed in the literature \cite{ref15,ref17}.

\begin{figure}[!t]
\centering
\begin{minipage}[b]{0.45\linewidth}
\includegraphics[width=\linewidth]{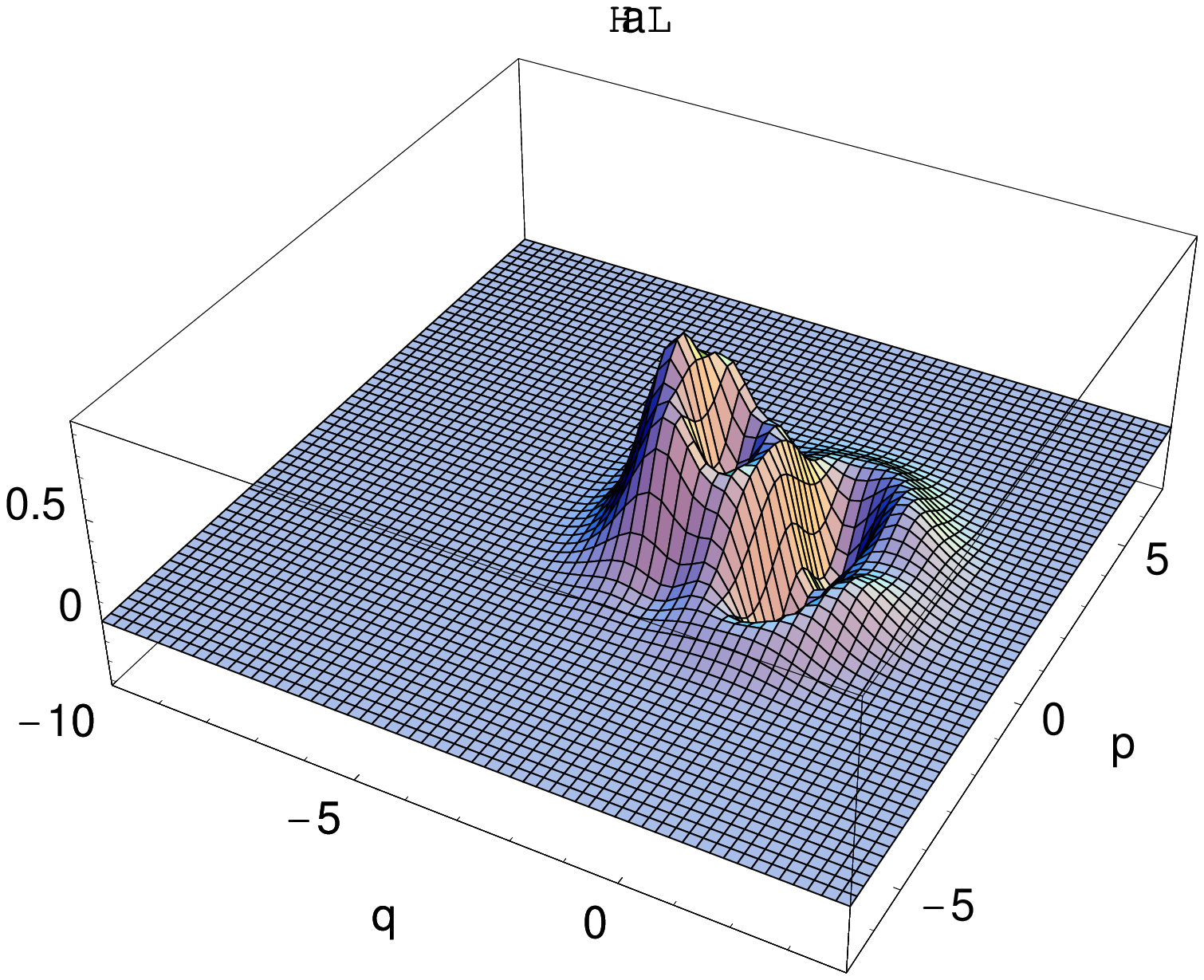}
\end{minipage} \hfill
\begin{minipage}[b]{0.45\linewidth}
\includegraphics[width=\linewidth]{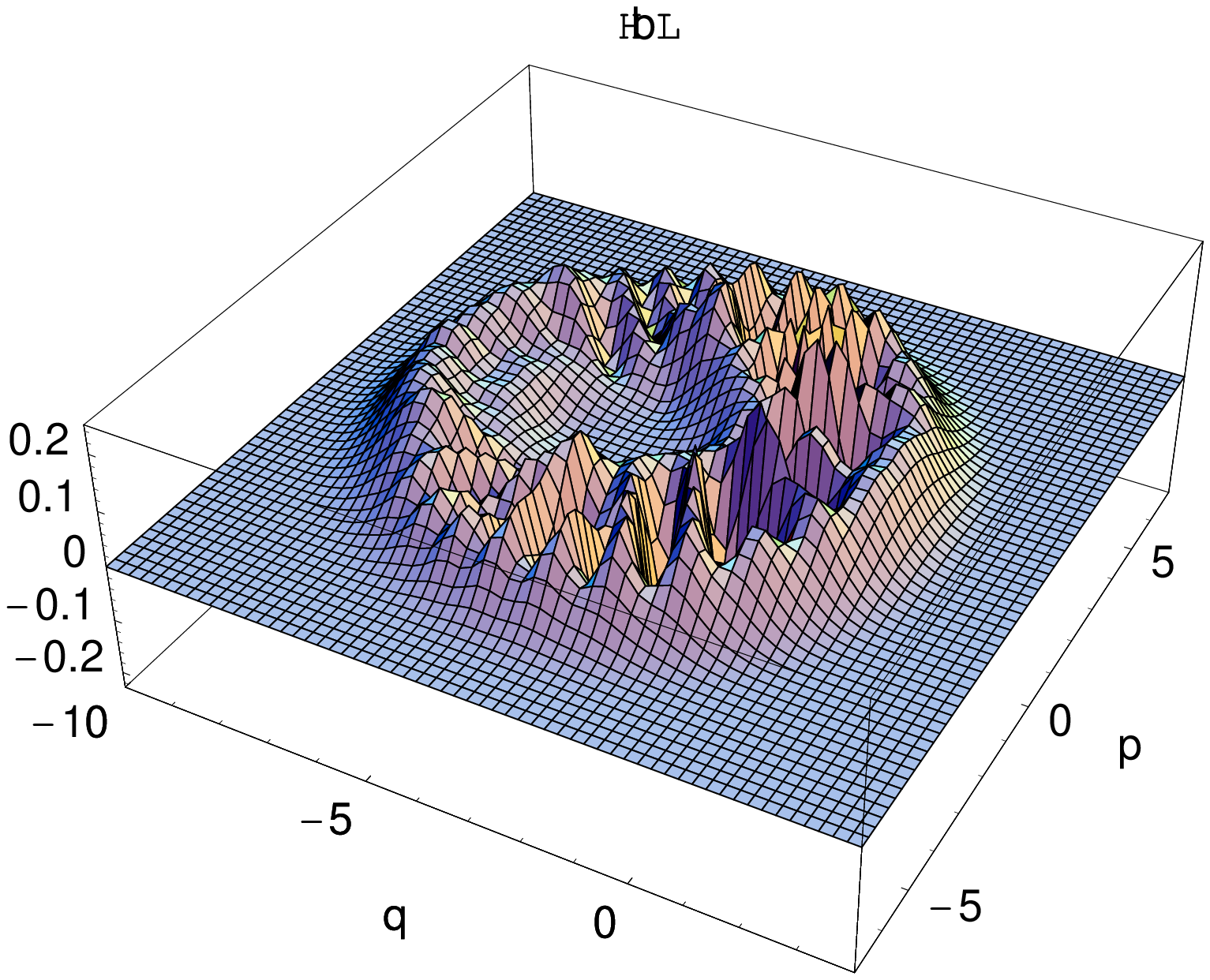}
\end{minipage} \hfill
\begin{minipage}[b]{0.45\linewidth}
\includegraphics[width=\linewidth]{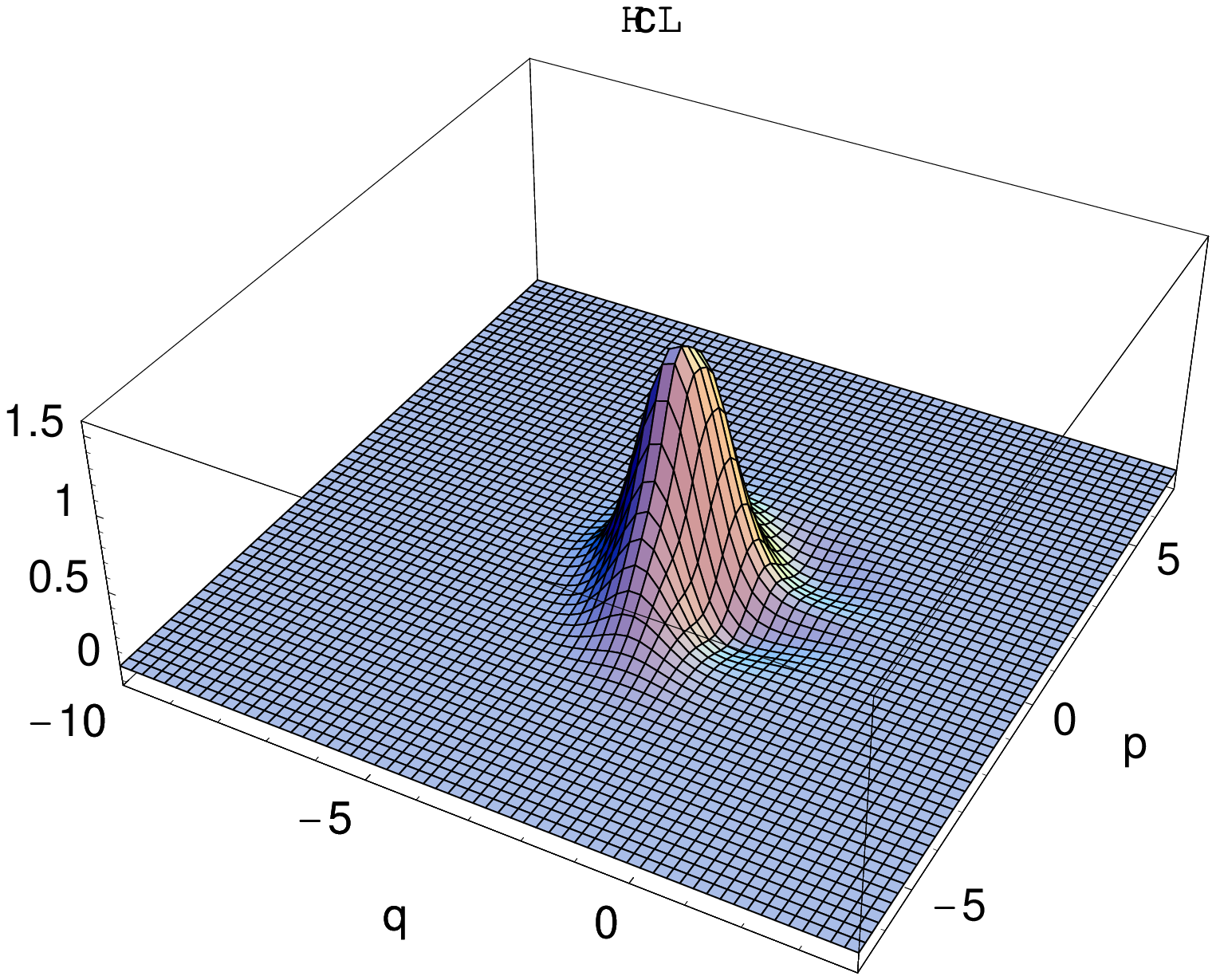}
\end{minipage} \hfill
\begin{minipage}[b]{0.45\linewidth}
\includegraphics[width=\linewidth]{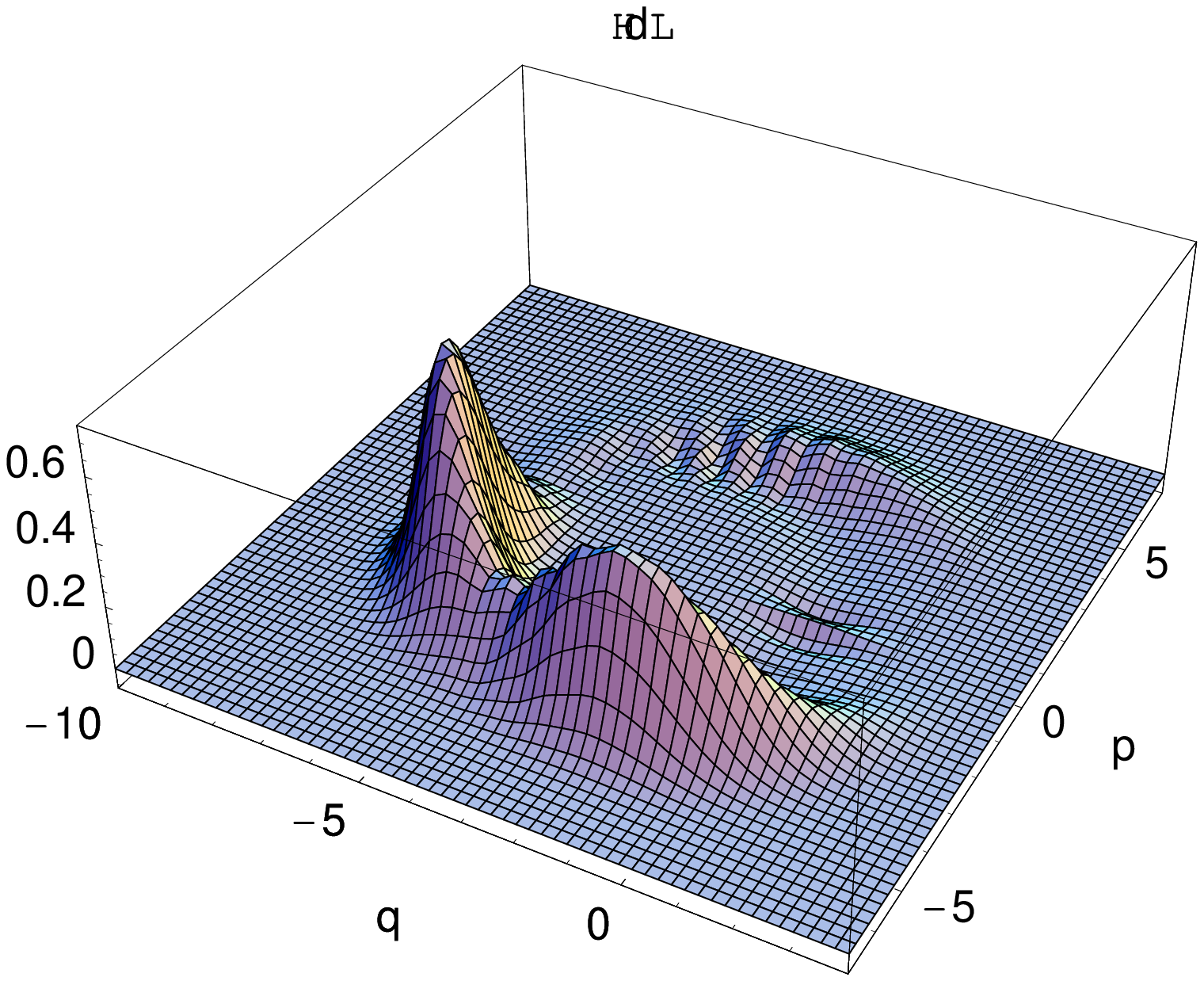}
\end{minipage}
\caption{The Wigner function $W_{{\rm a}}^{({\rm o})}(\gamma;t)$ is plotted assuming the same set of parameters
established in the previous figure for the detuning frequency and amplitude of the driving field, with $| \alf | = 1$
$(\lg {\bf n}_{{\rm a}} \rg_{{\rm o}} \approx 1.313)$ and $\kappa_{{\rm eff}} t = 100$ fixed. Note that the entanglement
is maximum when $\delta = 0$ (resonant regime), and minimum for $\delta = 10 \kappa_{{\rm eff}}$ (nonresonant regime).}
\end{figure}
For instance, let us consider the Glauber-Sudarshan quasiprobability distributions for even- and odd-coherent states into
the Eq. (\ref{e21}). After the integrations in the complex $\alf_{{\rm a}}$- and $\alf_{{\rm b}}$-planes, we get
\be
\lb{e23}
W_{{\rm a}}^{({\rm e})}(\gamma;t) = \frac{\exp (| \alf |^{2})}{4 \cosh (| \alf |^{2})} \lbk {\rm K}_{\gamma}(\alf,\beta;t)
+ {\rm K}_{\gamma}(-\alf,\beta;t) + \exp (-2 | \alf |^{2}) \mathbb{K}_{\gamma}(\alf,\beta;t) \rbk
\ee
and
\be
\lb{e24}
W_{{\rm a}}^{({\rm o})}(\gamma;t) = \frac{\exp (| \alf |^{2})}{4 \sinh (| \alf |^{2})} \lbk {\rm K}_{\gamma}(\alf,\beta;t)
+ {\rm K}_{\gamma}(-\alf,\beta;t) - \exp (-2 | \alf |^{2}) \mathbb{K}_{\gamma}(\alf,\beta;t) \rbk \; ,
\ee
where
\be
\lb{e25}
\mathbb{K}_{\gamma}(\alf,\beta;t) = 2 \exp (| \alf |^{2}) \left. \cosh \lpar \alf \frac{\partial}{\partial \alf_{{\rm a}}}
- \alf^{\ast} \frac{\partial}{\partial \alf_{{\rm a}}^{\ast}} \rpar \lbk \exp (| \alf_{{\rm a}} |^{2}) {\rm K}_{\gamma}
(\alf_{{\rm a}},\beta;t) \rbk \right|_{\alf_{{\rm a}} = 0} \; .
\ee
Note that at time $t=0$, the function $\exp (-2 |\alf|^{2}) \mathbb{K}_{\gamma}(\alf,\beta;0) = 4 \exp (-2 |\gamma|^{2})
\cos [4 \ima (\gamma \alf^{\ast})]$ leads us to recover well-known expressions in the literature \cite{ref29}:
\bd
W_{{\rm a}}^{({\rm e})}(\gamma;0) = \frac{\exp (| \alf |^{2})}{\cosh (| \alf |^{2})} \exp(-2 |\gamma|^{2}) \lbr \exp
(-2 |\alf|^{2}) \cosh [4 \re (\gamma \alf^{\ast})] + \cos [4 \ima (\gamma \alf^{\ast})] \rbr
\ed
and
\bd
W_{{\rm a}}^{({\rm o})}(\gamma;0) = \frac{\exp (| \alf |^{2})}{\sinh (| \alf |^{2})} \exp(-2 |\gamma|^{2}) \lbr \exp
(-2 |\alf|^{2}) \cosh [4 \re (\gamma \alf^{\ast})] - \cos [4 \ima (\gamma \alf^{\ast})] \rbr \; .
\ed
Figs. 4 and 5 show the three-dimensional plots of $W_{{\rm a}}^{({\rm e})}(\gamma;t)$ and $W_{{\rm a}}^{({\rm o})}
(\gamma;t)$ versus $p = \sqrt{2} \, \ima (\gamma)$ and $q = \sqrt{2} \, \re (\gamma)$, respectively, for the atom-cavity
system resonant (a,b) $\delta = 0$ and nonresonant (c,d) $\delta = 10 \kappa_{{\rm eff}}$. In both simulations, we consider
$| \alf | = 1$ and two different values of amplitude of the driving field: (a,c) $| \beta | = 2$ and (b,d) $| \beta | = 5$.
Furthermore, we also fix the parameter $\kappa_{{\rm eff}} t = 100$ which permits us to obtain a partial view of
entanglement in the tripartite system. A first analysis of these pictures shows, via Wigner function, that the entanglement
is sensitive to variations of the experimental parameters $| \beta |$ and $\delta$ (this fact corroborates the previous
results obtained for atomic inversion); being the detuning parameter responsible for the entanglement degree between the
components involved in the system, since both the driving and cavity fields are in resonance. In this sense, although the
Figs. 4(c)-(d) and 5(c)-(d) have similar structures, there are subtle differences between them: $W_{{\rm a}}^{({\rm o})}
(\gamma;t)$ assumes negative values due to initial sub-Poissonian photon statistics of the cavity field; while
$W_{{\rm a}}^{({\rm e})}(\gamma;t)$ is strictly positive, since the even-coherent state has super-Poissonian photon
statistics for any initial value of $\lg {\bf n}_{{\rm a}} \rg_{{\rm e}}$ (see Ref. \cite{ref29} for more details). On the
other hand, the increase of $| \beta |$ in Figs. 4(b,d) and 5(b,d) shows an interesting effect on the Wigner functions:
the interference patterns between the states of the driving and cavity fields turn to be more pronounced, and this effect
modifies the shapes of $W_{{\rm a}}^{({\rm e})}(\gamma;t)$ and $W_{{\rm a}}^{({\rm o})}(\gamma;t)$. Similar analysis can be
also applied if one considers both the external and cavity fields in the coherent states (see appendix B).

To conclude this section, we will determine the marginal probability distribution functions $|\psi_{{\rm a}} (q;t) |^{2}$ 
and $| \varphi_{{\rm a}} (p;t) |^{2}$ through the direct integration of Eq. (\ref{e21}) over the variables $p$ or $q$, i.e.,
\br
\lb{e26}
| \psi_{{\rm a}} (q;t) |^{2} &=& \int \!\!\!\!\! \int \frac{d^{2} \alf_{{\rm a}} d^{2} \alf_{{\rm b}}}{\pi^{2}} \, 
P_{{\rm a}}(\alf_{{\rm a}}) P_{{\rm b}}(\alf_{{\rm b}}) \, {\rm U}_{q} ( \alf_{{\rm a}},\alf_{{\rm b}};t ) \; , \\
\lb{e27}
| \varphi_{{\rm a}} (p;t) |^{2} &=& \int \!\!\!\!\! \int \frac{d^{2} \alf_{{\rm a}} d^{2} \alf_{{\rm b}}}{\pi^{2}} \, 
P_{{\rm a}}(\alf_{{\rm a}}) P_{{\rm b}}(\alf_{{\rm b}}) \, {\rm V}_{p} ( \alf_{{\rm a}},\alf_{{\rm b}};t ) \; , 
\er
with
\bd
{\rm U}_{q} ( \alf_{{\rm a}},\alf_{{\rm b}};t ) = \int_{- \infty}^{\infty} \frac{dp}{\sqrt{2 \pi}} \, {\rm K}_{\gamma} 
( \alf_{{\rm a}},\alf_{{\rm b}};t ) \qquad \mbox{and} \qquad {\rm V}_{p} ( \alf_{{\rm a}},\alf_{{\rm b}};t ) 
= \int_{- \infty}^{\infty} \frac{dq}{\sqrt{2 \pi}} \, {\rm K}_{\gamma}( \alf_{{\rm a}},\alf_{{\rm b}};t ) \; . 
\ed
For this purpose, let us initially introduce the complex function
\br
{\cal H}_{\mu}^{(m,m')}(\alf_{{\rm a}},\alf_{{\rm b}}) &=& \sum_{k=0}^{ \{ m,m' \} } (2k)!! \, L_{k}^{(m-k)}(0) \, 
L_{k}^{(m'-k)}(0) \lbk \frac{\eps_{{\rm a}}^{3} \alf_{{\rm a}} + \eps_{{\rm b}}^{3} \alf_{{\rm b}}}{\eps_{{\rm a}} 
\eps_{{\rm b}} (\eps_{{\rm b}} \alf_{{\rm a}} + \eps_{{\rm a}} \alf_{{\rm b}})} \rbk^{k} \nn \\
& & \times \, H_{m-k} \lpar \mu - \frac{\nu_{{\rm a}} + \nu_{{\rm b}}^{\ast}}{\sqrt{2}} \rpar H_{m'-k} \lpar \mu - 
\frac{\nu_{{\rm a}} + \nu_{{\rm b}}^{\ast}}{\sqrt{2}} \rpar \; , \nn
\er
where $H_{n}(z)$ is the Hermite polynomial, $\nu_{{\rm a}({\rm b})} = \eps_{{\rm a}({\rm b})} \lpar \eps_{{\rm a}({\rm b})} 
\alf_{{\rm a}} - \eps_{{\rm b}({\rm a})} \alf_{{\rm b}} \rpar$, and $\{ m,m' \}$ stands for the minor of $m$ and $m'$. In 
addition, we define the auxiliary functions
\bd
{\rm Y}_{\mu}^{(m,m')}(\alf_{{\rm a}},\alf_{{\rm b}};t) = {\cal H}_{\mu}^{(m,m')}(\alf_{{\rm a}},\alf_{{\rm b}}) \, 
F_{m}(t) F_{m'}^{\ast}(t) + \frac{\eps_{{\rm a}} \eps_{{\rm b}}}{2} \frac{\eps_{{\rm b}} \alf_{{\rm a}} + \eps_{{\rm a}} 
\alf_{{\rm b}}}{\eps_{{\rm a}} \alf_{{\rm a}} + \eps_{{\rm b}} \alf_{{\rm b}}} \, {\cal H}_{\mu}^{(m+1,m'+1)}
(\alf_{{\rm a}},\alf_{{\rm b}}) \, \frac{G_{m}(t) G_{m'}^{\ast}(t)}{\sqrt{(m+1)(m'+1)}}
\ed
and 
\bd
{\cal A}_{\mu}^{(m,m')}(\alf_{{\rm a}},\alf_{{\rm b}}) = \sqrt{2} \, \exp \lbk - \lpar \mu - \frac{\nu_{{\rm a}} + 
\nu_{{\rm b}}^{\ast}}{\sqrt{2}} \rpar^{2} - \left| \eps_{{\rm a}} \alf_{{\rm a}} + \eps_{{\rm b}} \alf_{{\rm b}} \right|^{2}
\rbk \frac{\lbk \sqrt{2} \, \eps_{{\rm b}} \lpar \eps_{{\rm b}} \alf_{{\rm a}} + \eps_{{\rm a}} \alf_{{\rm b}} \rpar 
\rbk^{m} \lbk \sqrt{2} \, \eps_{{\rm a}} \lpar \eps_{{\rm a}} \alf_{{\rm a}} + \eps_{{\rm b}} \alf_{{\rm b}} \rpar^{\ast} 
\rbk^{m'}}{(2m)!! \, (2m')!!} \; , 
\ed
which permit us to express the integrands ${\rm U}_{q}(\alf_{{\rm a}},\alf_{{\rm b}};t)$ and ${\rm V}_{p}(\alf_{{\rm a}},
\alf_{{\rm b}};t)$ in compact forms as follows:
\br
\lb{e28}
{\rm U}_{q}(\alf_{{\rm a}},\alf_{{\rm b}};t) &=& \sum_{m,m'=0}^{\infty} {\cal A}_{q}^{(m,m')}(\alf_{{\rm a}},\alf_{{\rm b}})
\, {\rm Y}_{q}^{(m,m')}(\alf_{{\rm a}},\alf_{{\rm b}};t) \; , \\
\lb{e29}
{\rm V}_{p}(\alf_{{\rm a}},\alf_{{\rm b}};t) &=& \sum_{m,m'=0}^{\infty} {\cal A}_{p}^{(m,m')}(-\im \alf_{{\rm a}},- \im
\alf_{{\rm b}}) \, {\rm Y}_{p}^{(m,m')}(- \im \alf_{{\rm a}},- \im \alf_{{\rm b}};t) \; .
\er
Consequently, the connection between Eqs. (\ref{e28}) and (\ref{e29}) can be promptly established through the mathematical
relations ${\rm U}_{q}(\alf_{{\rm a}},\alf_{{\rm b}};t) = {\rm V}_{q}(\im \alf_{{\rm a}}, \im \alf_{{\rm b}};t)$ and 
${\rm V}_{p}(\alf_{{\rm a}},\alf_{{\rm b}};t) = {\rm U}_{p}(- \im \alf_{{\rm a}},- \im \alf_{{\rm b}};t)$. 

In analogy to Wigner function, the marginal probability distribution functions do not depend on the driving field at time 
$t=0$, since their expressions are reduced to 
\br
| \psi_{{\rm a}} (q;0) |^{2} &=& \sqrt{2} \int \frac{d^{2} \alf_{{\rm a}}}{\pi} \exp \{ - [q - \sqrt{2} \, \re 
(\alf_{{\rm a}}) ]^{2} \} \, P_{{\rm a}}(\alf_{{\rm a}}) \; , \nn \\
| \varphi_{{\rm a}} (p;0) |^{2} &=& \sqrt{2} \int \frac{d^{2} \alf_{{\rm a}}}{\pi} \exp \{ - [p - \sqrt{2} \, \ima 
(\alf_{{\rm a}}) ]^{2} \} \, P_{{\rm a}}(\alf_{{\rm a}}) \; . \nn
\er
Now, if one considers both the external and cavity fields in the coherent states, we get $| \psi_{{\rm a}} (q;t) |^{2} = 
{\rm U}_{q}(\alf,\beta;t)$ and $| \varphi_{{\rm a}} (p;t) |^{2} = {\rm V}_{p}(\alf,\beta;t)$. In particular, this example 
shows that the marginal distributions represent an important additional tool in the qualitative study of entanglement, since 
the variables $\alf$ and $\beta$ are completely correlated.

\section{Summary and conclusions}

In this paper, we have applied the decomposition formula for ${{\rm SU}}(2)$ Lie algebra on the driven Jaynes-Cummings
model in order to calculate, for instance, the exact expressions for atomic inversion and Wigner function when the atom is 
initially prepared at the excited state. In fact, adopting the diagonal representation of coherent states, we have shown 
that these expressions can be written in the integral form, with their integrands presenting a commom term which describes 
the product of the Glauber-Sudarshan quasiprobability distribution functions for each field, and a kernel responsible for 
the entanglement. It is important mentioning that the mathematical procedure developed here does not present any 
restrictions on the states of the cavity and driving electromagnetic fields. Following, to illustrate these results we have 
fixed the driving field in the coherent state and assumed two different possibilities for the cavity field (i.e., the even- 
and odd-coherent states). In this way, we have verified that the amplitude of the external field and detuning parameter 
(i) perform a strong influence on the structures of collapses and revivals in the atomic inversion, (ii) control the 
entanglement degree in the tripartite system; and consequently, (iii) modify the shape of $W_{{\rm a}}(\gamma;t)$ since the 
interference patterns between the states of the driving and cavity fields turn to be more evident through the Wigner 
function. In addition, the formalism employed in the calculation of atomic inversion and Wigner function open new
possibilities of future investigations in similar physical systems (e.g., see Refs. \cite{ref26,ref27}); or in the study
of dissipative composite systems, where the decoherence effect has a central role in the quantum information processing.
These considerations are under current research and will be published elsewhere. Summarizing, the work reported here is 
clearly the product of considerable effort and represents an original contribution to the wider field of entangled-state 
engineering with emphasis on quantum computation and related topics.

\section*{Acknowledgments}

The authors are grateful to R.J. Napolitano and V. V. Dodonov for reading the manuscript and for providing valuable suggestions. MAM
acknowledges financial support from FAPESP, S\~{a}o Paulo, Brazil, project no. 01/11209-0. RJM and JAR acknowledge
financial support from CAPES and CNPq, respectively, both Brazilian agencies. This work was supported by FAPESP through the
project no. 00/15084-5, and it is also linked to the Optics and Photonics Research Center.

\appendix
\section{The integral form of the atomic inversion}

With the help of the definition established in Sec. III for atomic inversion and the cyclic invariance property of the
trace operation, we get
\br
\lb{a1}
{\cal I}(t) &=& \tr_{{\rm ab}} \lbr \ro_{{\rm ab}}(0) \lbk \bu_{11}^{\dagger}(t) \bu_{11}(t) - \bu_{21}^{\dagger}(t)
\bu_{21}(t) \rbk \rbr \nn \\
&=& \tr_{{\rm ab}} \lbr \ro_{{\rm ab}}(0) \lbk \cos \lpar 2 t \sqrt{\bet_{{\rm A}}} \; \rpar + \frac{\delta^{2}}{2}
\frac{\sin^{2} \lpar t \sqrt{\bet_{{\rm A}}} \; \rpar}{\bet_{{\rm A}}} \rbk \rbr \; .
\er
Employing the diagonal representation of $\ro_{{\rm ab}}(0)$ in the coherent states basis into the second equality of
Eq. (\ref{a1}), the integral form of the atomic inversion can be promptly obtained, i.e.,
\be
\lb{a2}
{\cal I}(t) = \int \!\!\!\!\! \int \frac{d^{2} \alf_{{\rm a}} d^{2} \alf_{{\rm b}}}{\pi^{2}} \, P_{{\rm a}}(\alf_{{\rm a}})
P_{{\rm b}}(\alf_{{\rm b}}) \, \Xi ( \alf_{{\rm a}},\alf_{{\rm b}};t )
\ee
where
\be
\lb{a3}
\Xi ( \alf_{{\rm a}},\alf_{{\rm b}};t ) = \lg \alf_{{\rm a}},\alf_{{\rm b}} | \cos ( 2t \sqrt{\bet_{{\rm A}}} \, ) |
\alf_{{\rm a}},\alf_{{\rm b}} \rg + \frac{\delta^{2}}{2} \, \lg \alf_{{\rm a}},\alf_{{\rm b}} | \frac{\sin^{2} ( t
\sqrt{\bet_{{\rm A}}} \, )}{\bet_{{\rm A}}} | \alf_{{\rm a}},\alf_{{\rm b}} \rg \; .
\ee
However, the effectiveness of the integral form (\ref{a2}) is connected with the determination of an analytical expression
for Eq. (\ref{a3}).

To calculate the function $\Xi (\alf_{{\rm a}},\alf_{{\rm b}};t)$, firstly we expand the operators $\cos ( 2t
\sqrt{\bet_{{\rm A}}} \, )$ and $\sin^{2} ( t \sqrt{\bet_{{\rm A}}} \, ) / \bet_{{\rm A}}$ in a power series as follows:
\bd
\cos ( 2t \sqrt{\bet_{{\rm A}}} \, ) = \left. \sum_{k=0}^{\infty} \frac{(-1)^{k}}{(2k)!} ( \sqrt{2} \kappa_{{\rm eff}}
t )^{2k} \frac{d^{k}}{dx^{k}} \, e^{2x ( 1 + \delta^{2} / 4 \kappa_{{\rm eff}}^{2} )} \, e^{x {\bf D}} \, e^{x {\bf S}}
\right|_{x=0} \; ,
\ed
and
\bd
\frac{\sin^{2} ( t \sqrt{\bet_{{\rm A}}} \, )}{\bet_{{\rm A}}} = \frac{1}{\kappa_{{\rm eff}}^{2}} \left. \sum_{k=0}^{\infty}
\frac{(-1)^{k}}{[2(k+1)]!} ( \sqrt{2} \kappa_{{\rm eff}} t )^{2(k+1)} \frac{d^{k}}{dx^{k}} \, e^{2x ( 1 + \delta^{2} / 4
\kappa_{{\rm eff}}^{2} )} \, e^{x {\bf D}} \, e^{x {\bf S}} \right|_{x=0} \; .
\ed
Secondly, we apply the antinormal-order decomposition formula for ${{\rm SU}}(2)$ Lie algebra on the operator
$e^{x {\bf D}}$, which leads us to obtain \cite{ap1,ap2,ap3}
\bd
e^{x {\bf D}} = e^{B_{+} {\bf K}_{-}} \, e^{B_{+} B_{0} {\bf K}_{+}} \, e^{(\ln B_{0}) {\bf K}_{0}} \; ,
\ed
where
\bd
B_{+} = \frac{2 \eps_{{\rm a}} \eps_{{\rm b}} \sinh x}{\cosh x + (\eps_{{\rm a}}^{2} - \eps_{{\rm b}}^{2}) \sinh x} \qquad
\mbox{and} \qquad B_{0} = \lbk \cosh x + (\eps_{{\rm a}}^{2} - \eps_{{\rm b}}^{2}) \sinh x \rbk^{2} \; .
\ed
After lengthy calculations, the analytical expressions for the mean values
\be
\lb{a4}
\lg \alf_{{\rm a}},\alf_{{\rm b}} | \cos ( 2t \sqrt{\bet_{{\rm A}}} \, ) | \alf_{{\rm a}},\alf_{{\rm b}} \rg = \exp \lpar
- \left| \eps_{{\rm a}} \alf_{{\rm a}} + \eps_{{\rm b}} \alf_{{\rm b}} \right|^{2} \rpar \sum_{n=0}^{\infty} \frac{\left|
\eps_{{\rm a}} \alf_{{\rm a}} + \eps_{{\rm b}} \alf_{{\rm b}} \right|^{2n}}{n!} \cos (t \Delta_{n})
\ee
and
\be
\lb{a5}
\lg \alf_{{\rm a}},\alf_{{\rm b}} | \frac{\sin^{2} ( t \sqrt{\bet_{{\rm A}}} \, )}{\bet_{{\rm A}}} | \alf_{{\rm a}},
\alf_{{\rm b}} \rg = \exp \lpar - \left| \eps_{{\rm a}} \alf_{{\rm a}} + \eps_{{\rm b}} \alf_{{\rm b}} \right|^{2} \rpar
\sum_{n=0}^{\infty} \frac{\left| \eps_{{\rm a}} \alf_{{\rm a}} + \eps_{{\rm b}} \alf_{{\rm b}} \right|^{2n}}{n!}
\frac{\sin^{2} (t \Delta_{n} /2)}{(\Delta_{n} /2)^{2}}
\ee
are determined, with $\Delta_{n}^{2} = \delta^{2} + \Omega_{n}^{2}$ and $\Omega_{n} = 2 \kappa_{{\rm eff}} \sqrt{n+1}$
the effective Rabi frequency. Now, substituting these results into Eq. (\ref{a3}), we obtain
\be
\lb{a6}
\Xi (\alf_{{\rm a}},\alf_{{\rm b}};t) = 1 - 2 \exp \lpar - \left| \eps_{{\rm a}} \alf_{{\rm a}} + \eps_{{\rm b}}
\alf_{{\rm b}} \right|^{2} \rpar \sum_{n=0}^{\infty} \frac{\left| \eps_{{\rm a}} \alf_{{\rm a}} + \eps_{{\rm b}}
\alf_{{\rm b}} \right|^{2n}}{n!} |G_{n}(t)|^{2} \; ,
\ee
where $G_{n}(t) = - \im (\Omega_{n} / \Delta_{n}) \sin (\Delta_{n} t/2)$. Consequently, with the determination of the
analytical expression for $\Xi (\alf_{{\rm a}},\alf_{{\rm b}};t)$, the effectiveness of the integral form (\ref{a2}) is
guaranteed.

\section{Calculational details of the Wigner function}

Initially, we will derive the means values
\be
\lb{b1}
\lg \alf_{{\rm a}},\alf_{{\rm b}} | \bu_{11}^{\dagger}(t) {\bf D}_{{\rm a}}(\xi) \bu_{11}(t) | \alf_{{\rm a}},\alf_{{\rm b}}
\rg = \int \!\!\!\!\! \int \frac{d^{2} \beta_{{\rm a}} d^{2} \beta_{{\rm b}}}{\pi^{2}} \, \lg \alf_{{\rm a}},\alf_{{\rm b}}
| \bu_{11}^{\dagger}(t) {\bf D}_{{\rm a}}(\xi) | \beta_{{\rm a}},\beta_{{\rm b}} \rg \lg \beta_{{\rm a}},\beta_{{\rm b}} |
\bu_{11}(t) | \alf_{{\rm a}},\alf_{{\rm b}} \rg
\ee
and
\be
\lb{b2}
\lg \alf_{{\rm a}},\alf_{{\rm b}} | \bu_{21}^{\dagger}(t) {\bf D}_{{\rm a}}(\xi) \bu_{21}(t) | \alf_{{\rm a}},\alf_{{\rm b}}
\rg = \int \!\!\!\!\! \int \frac{d^{2} \beta_{{\rm a}} d^{2} \beta_{{\rm b}}}{\pi^{2}} \, \lg \alf_{{\rm a}},\alf_{{\rm b}}
| \bu_{21}^{\dagger}(t) {\bf D}_{{\rm a}}(\xi) | \beta_{{\rm a}},\beta_{{\rm b}} \rg \lg \beta_{{\rm a}},\beta_{{\rm b}} |
\bu_{21}(t) | \alf_{{\rm a}},\alf_{{\rm b}} \rg \; ,
\ee
by means of integrations in the complex variables $\beta_{{\rm a}}$ and $\beta_{{\rm b}}$. Thus, let us substitute into
Eqs. (\ref{b1}) and (\ref{b2}) the auxiliary mean values
\br
\lg \alf_{{\rm a}},\alf_{{\rm b}} | \bu_{11}^{\dagger}(t) {\bf D}_{{\rm a}}(\xi) | \beta_{{\rm a}},\beta_{{\rm b}} \rg
&=& \exp \lbk \half \lpar \xi \beta_{{\rm a}}^{\ast} - \xi^{\ast} \beta_{{\rm a}} \rpar \rbk \lpar \lg \beta_{{\rm a}} +
\xi,\beta_{{\rm b}} | \bu_{11}(t) | \alf_{{\rm a}},\alf_{{\rm b}} \rg \rpar^{\ast} \; , \nn \\
\lg \alf_{{\rm a}},\alf_{{\rm b}} | \bu_{21}^{\dagger}(t) {\bf D}_{{\rm a}}(\xi) | \beta_{{\rm a}},\beta_{{\rm b}} \rg
&=& \exp \lbk \half \lpar \xi \beta_{{\rm a}}^{\ast} - \xi^{\ast} \beta_{{\rm a}} \rpar \rbk \lpar \lg \beta_{{\rm a}} +
\xi,\beta_{{\rm b}} | \bu_{21}(t) | \alf_{{\rm a}},\alf_{{\rm b}} \rg \rpar^{\ast} \; , \nn \\
\lg \beta_{{\rm a}},\beta_{{\rm b}} | \bu_{11}(t) | \alf_{{\rm a}},\alf_{{\rm b}} \rg &=& \sum_{m=0}^{\infty}
F_{m}(t) \, \Lambda_{m}(\alf_{{\rm a}},\alf_{{\rm b}},\beta_{{\rm a}},\beta_{{\rm b}}) \; , \nn \\
\lg \beta_{{\rm a}},\beta_{{\rm b}} | \bu_{21}(t) | \alf_{{\rm a}},\alf_{{\rm b}} \rg &=& \lpar \eps_{{\rm a}}
\beta_{{\rm a}} + \eps_{{\rm b}} \beta_{{\rm b}} \rpar^{\ast} \sum_{m=0}^{\infty} \frac{G_{m}(t)}{\sqrt{m+1}} \,
\Lambda_{m}(\alf_{{\rm a}},\alf_{{\rm b}},\beta_{{\rm a}},\beta_{{\rm b}}) \; , \nn
\er
with
\br
\Lambda_{m}(\alf_{{\rm a}},\alf_{{\rm b}},\beta_{{\rm a}},\beta_{{\rm b}}) &=& \exp \lbk - \half \lpar |\alf_{{\rm a}}|^{2}
+ |\alf_{{\rm b}}|^{2} + |\beta_{{\rm a}}|^{2} + |\beta_{{\rm b}}|^{2} \rpar + \lpar \eps_{{\rm b}} \alf_{{\rm a}} -
\eps_{{\rm a}} \alf_{{\rm b}} \rpar \lpar \eps_{{\rm b}} \beta_{{\rm a}} - \eps_{{\rm a}} \beta_{{\rm b}} \rpar^{\ast} \rbk
\nn \\
& & \times \, \frac{\lbk \lpar \eps_{{\rm a}} \alf_{{\rm a}} + \eps_{{\rm b}} \alf_{{\rm b}} \rpar \lpar \eps_{{\rm a}}
\beta_{{\rm a}} + \eps_{{\rm b}} \beta_{{\rm b}} \rpar^{\ast} \rbk^{m}}{m!} \; , \nn
\er
and $F_{m}(t) = \cos (\Delta_{m} t/2) - \im (\delta / \Delta_{m}) \sin (\Delta_{m} t/2)$ (the function $G_{m}(t)$ was
previously defined in appendix A). Then, carrying out the integrations in the variables $\beta_{{\rm a}}$ and
$\beta_{{\rm b}}$, we get
\be
\lb{b3}
\lg \alf_{{\rm a}},\alf_{{\rm b}} | \bu_{11}^{\dagger}(t) {\bf D}_{{\rm a}}(\xi) \bu_{11}(t) | \alf_{{\rm a}},\alf_{{\rm b}}
\rg = \sum_{m,m'=0}^{\infty} \Upsilon_{\xi}^{(m,m')} (\alf_{{\rm a}},\alf_{{\rm b}}) {\rm I}_{\xi}^{(m,m')}(\alf_{{\rm a}},
\alf_{{\rm b}};t)
\ee
and
\be
\lb{b4}
\lg \alf_{{\rm a}},\alf_{{\rm b}} | \bu_{21}^{\dagger}(t) {\bf D}_{{\rm a}}(\xi) \bu_{21}(t) | \alf_{{\rm a}},\alf_{{\rm b}}
\rg = \sum_{m,m'=0}^{\infty} \Upsilon_{\xi}^{(m,m')} (\alf_{{\rm a}},\alf_{{\rm b}}) {\rm J}_{\xi}^{(m,m')}(\alf_{{\rm a}},
\alf_{{\rm b}};t) \; ,
\ee
where
\br
\Upsilon_{\xi}^{(m,m')} (\alf_{{\rm a}},\alf_{{\rm b}}) &=& \exp \lbk - \frac{| \xi |^{2}}{2} - \left| \eps_{{\rm a}}
\alf_{{\rm a}} + \eps_{{\rm b}} \alf_{{\rm b}} \right|^{2} + \eps_{{\rm b}} \lpar \eps_{{\rm b}} \alf_{{\rm a}} -
\eps_{{\rm a}} \alf_{{\rm b}} \rpar^{\ast} \xi - \eps_{{\rm a}} \lpar \eps_{{\rm a}} \alf_{{\rm a}} - \eps_{{\rm b}}
\alf_{{\rm b}} \rpar \xi^{\ast} \rbk \nn \\
& & \times \lbk \eps_{{\rm a}} \lpar \eps_{{\rm a}} \alf_{{\rm a}} + \eps_{{\rm b}} \alf_{{\rm b}} \rpar^{\ast} \xi
\rbk^{m'-m} \frac{\left| \eps_{{\rm a}} \alf_{{\rm a}} + \eps_{{\rm b}} \alf_{{\rm b}} \right|^{2m}}{m'!} \; , \nn \\
{\rm I}_{\xi}^{(m,m')}(\alf_{{\rm a}},\alf_{{\rm b}};t) &=& L_{m}^{(m'-m)} \lbk \eps_{{\rm a}} \eps_{{\rm b}}
\frac{\eps_{{\rm b}} \alf_{{\rm a}} + \eps_{{\rm a}} \alf_{{\rm b}}}{\eps_{{\rm a}} \alf_{{\rm a}} + \eps_{{\rm b}}
\alf_{{\rm b}}} | \xi |^{2} \rbk F_{m}(t) F_{m'}^{\ast}(t) \; , \nn \\
{\rm J}_{\xi}^{(m,m')}(\alf_{{\rm a}},\alf_{{\rm b}};t) &=& \sqrt{\frac{m+1}{m'+1}} \, L_{m+1}^{(m'-m)} \lbk \eps_{{\rm a}}
\eps_{{\rm b}} \frac{\eps_{{\rm b}} \alf_{{\rm a}} + \eps_{{\rm a}} \alf_{{\rm b}}}{\eps_{{\rm a}} \alf_{{\rm a}} +
\eps_{{\rm b}} \alf_{{\rm b}}} | \xi |^{2} \rbk G_{m}(t) G_{m'}^{\ast}(t) \; . \nn
\er
Consequently, the function $\widetilde{{\rm K}}_{\xi}(\alf_{{\rm a}},\alf_{{\rm b}};t)$ which appears in the integrand of
$\chi (\xi;t)$ can be determined as follows:
\be
\lb{b5}
\widetilde{{\rm K}}_{\xi}(\alf_{{\rm a}},\alf_{{\rm b}};t) = \sum_{m,m'=0}^{\infty} \Upsilon_{\xi}^{(m,m')}(\alf_{{\rm a}},
\alf_{{\rm b}}) \Gamma_{\xi}^{(m,m')}(\alf_{{\rm a}},\alf_{{\rm b}};t) \; ,
\ee
being $\Gamma_{\xi}^{(m,m')}(\alf_{{\rm a}},\alf_{{\rm b}};t) = {\rm I}_{\xi}^{(m,m')}(\alf_{{\rm a}},\alf_{{\rm b}};t) +
{\rm J}_{\xi}^{(m,m')}(\alf_{{\rm a}},\alf_{{\rm b}};t)$.

\begin{figure}[!t]
\centering
\begin{minipage}[b]{0.45\linewidth}
\includegraphics[width=\linewidth]{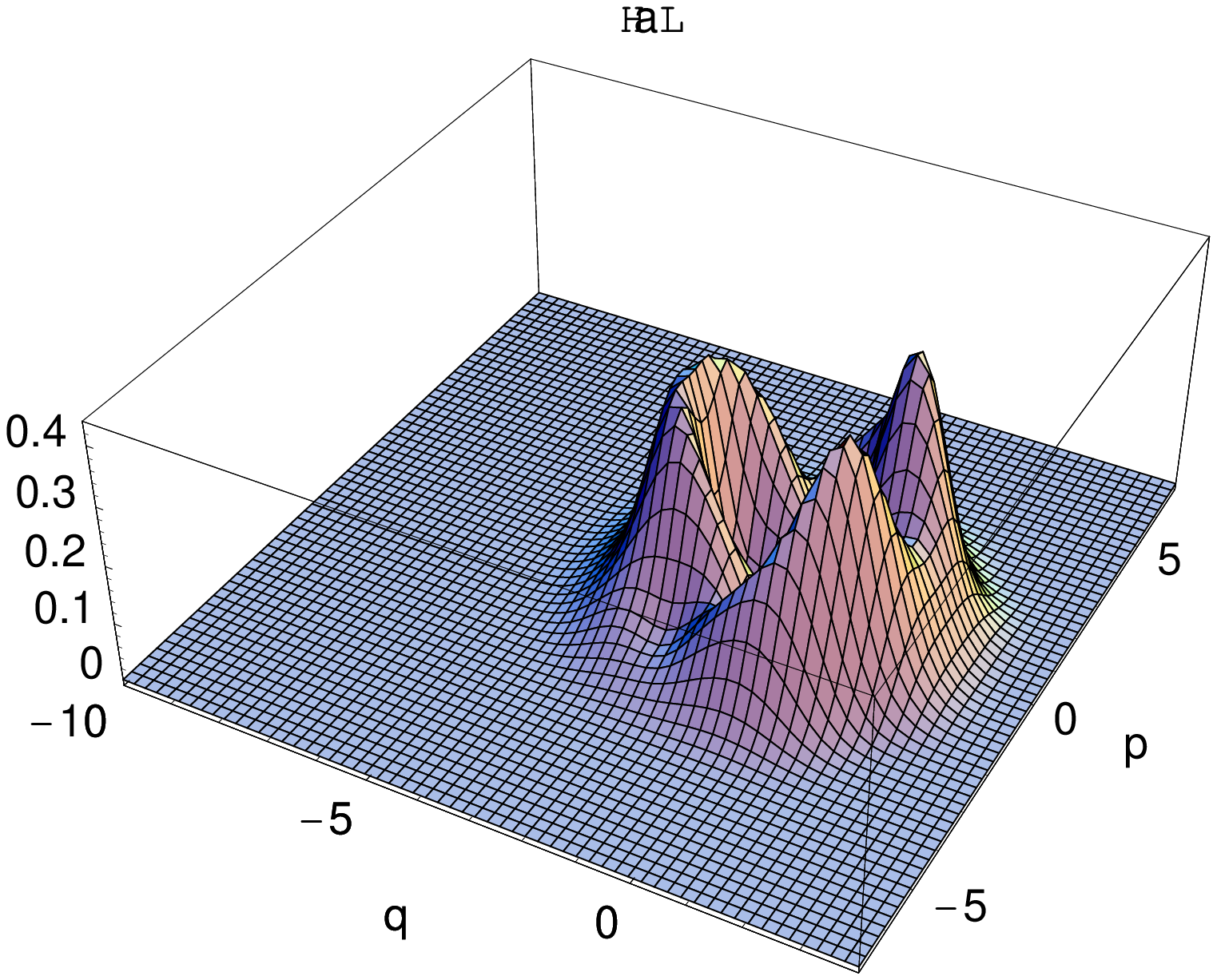}
\end{minipage} \hfill
\begin{minipage}[b]{0.45\linewidth}
\includegraphics[width=\linewidth]{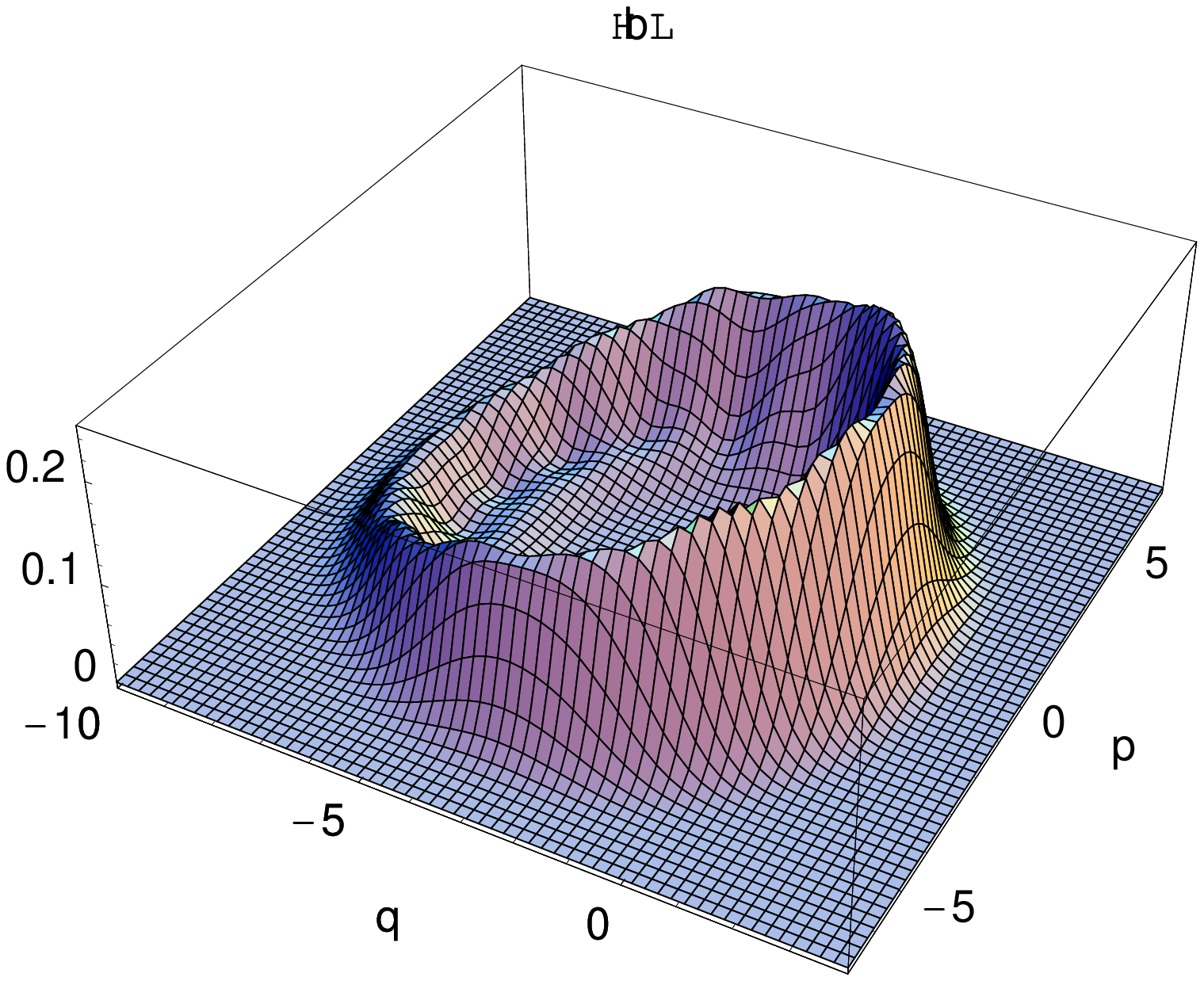}
\end{minipage} \hfill
\begin{minipage}[b]{0.45\linewidth}
\includegraphics[width=\linewidth]{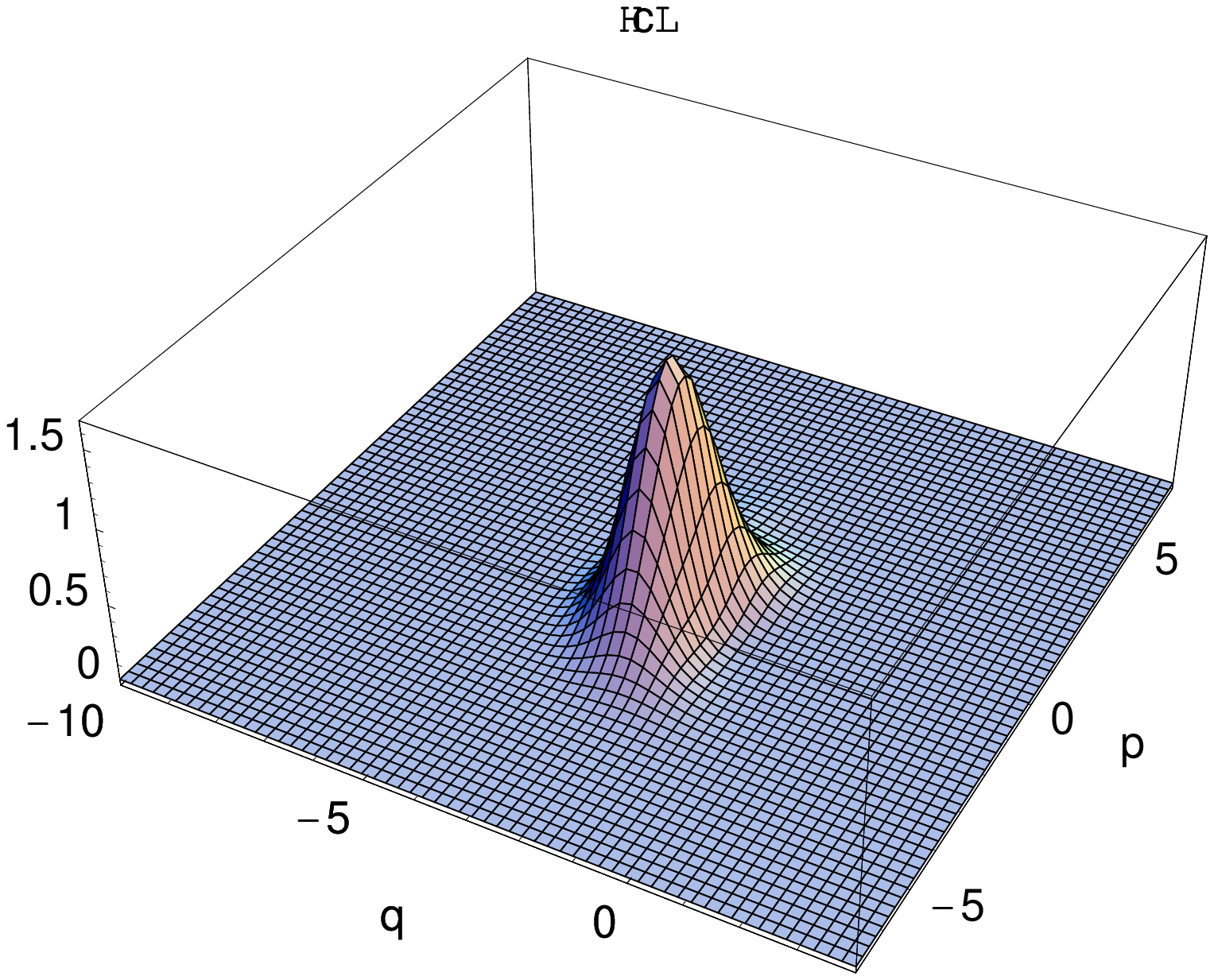}
\end{minipage} \hfill
\begin{minipage}[b]{0.45\linewidth}
\includegraphics[width=\linewidth]{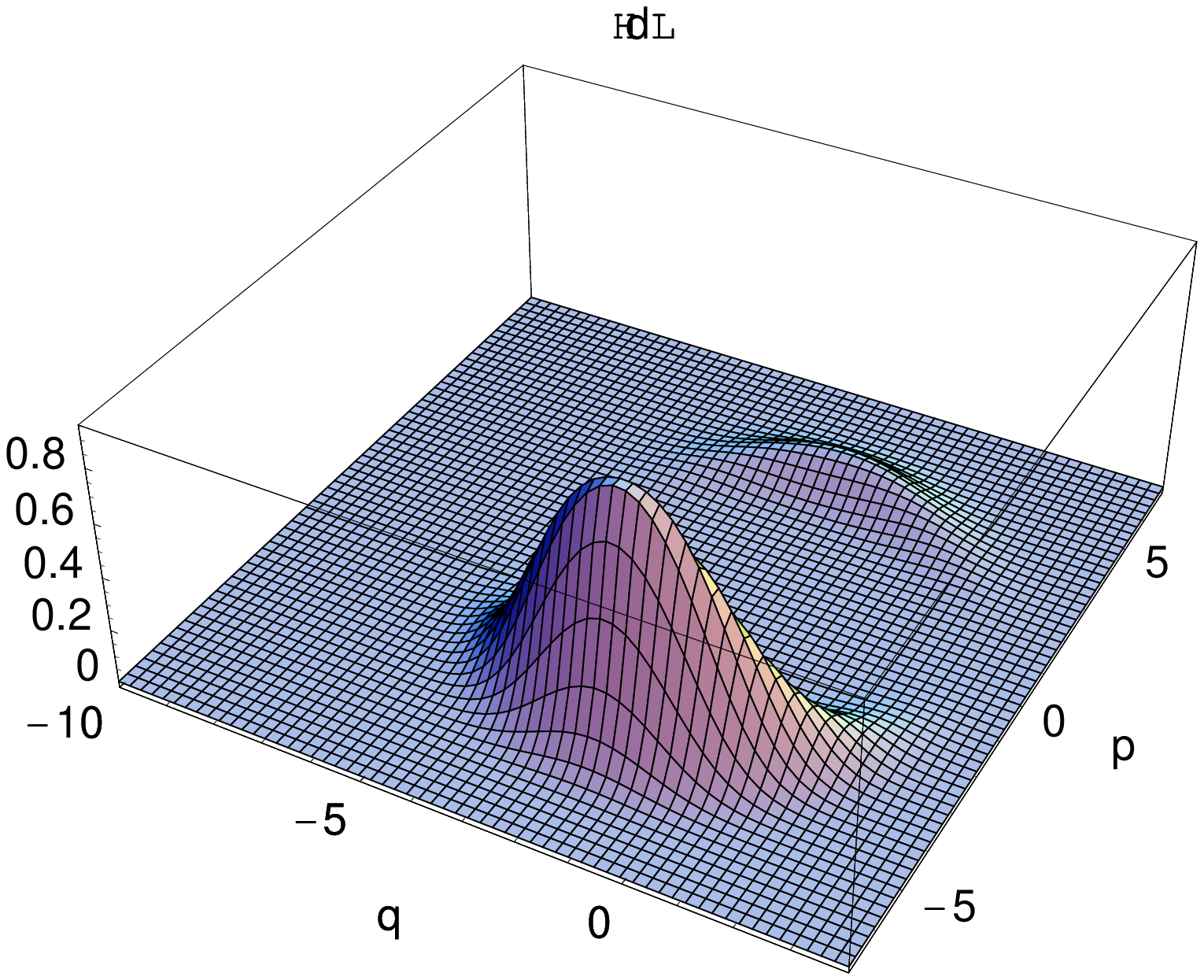}
\end{minipage}
\caption{Plots of $W_{{\rm a}}^{({\rm c})}(\gamma;t) = {\rm K}_{\gamma}(\alf,\beta;t)$ versus $p \in [-7,7]$ and
$q \in [-10,4]$ for the atom-cavity system resonant (a,b) $\delta = 0$ (maximum entanglement) and nonresonant (c,d)
$\delta = 10 \kappa_{{\rm eff}}$ (minimum entanglement), with $| \alf | = 1$ and $\kappa_{{\rm eff}} t = 100$ fixed. We
also have considered two different values of amplitude of the driving field: (a,c) $| \beta | = 2$ and (b,d)
$| \beta | = 5$, where the condition $\kappa_{{\rm a}({\rm b})} = \kappa$ was established in both situations.}
\end{figure}
An immediate application of this result is the calculation of ${\rm K}_{\gamma}(\alf_{{\rm a}},\alf_{{\rm b}};t)$ since
both functions are connected by a two-dimensional Fourier transform. Now, substituting (\ref{b5}) into Eq. (\ref{e22}) and
integrating in the complex variable $\xi$, we obtain as result the analytical expression
\be
\lb{b6}
{\rm K}_{\gamma}(\alf_{{\rm a}},\alf_{{\rm b}};t) = \sum_{m,m'=0}^{\infty} {\rm C}_{\gamma}^{(m,m')}(\alf_{{\rm a}},
\alf_{{\rm b}}) {\rm M}_{\gamma}^{(m,m')}(\alf_{{\rm a}},\alf_{{\rm b}};t) \; ,
\ee
where
\br
{\rm C}_{\gamma}^{(m,m')}(\alf_{{\rm a}},\alf_{{\rm b}}) &=& 2 \exp \lpar - \left| \eps_{{\rm a}} \alf_{{\rm a}} +
\eps_{{\rm b}} \alf_{{\rm b}} \right|^{2} - 2 \gamma_{{\rm a}} \gamma_{{\rm b}}^{\ast} \rpar \lbk 2 \eps_{{\rm a}} \lpar
\eps_{{\rm a}} \alf_{{\rm a}} + \eps_{{\rm b}} \alf_{{\rm b}} \rpar^{\ast} \gamma_{{\rm a}} \rbk^{m'-m} \nn \\
& & \times \, \frac{\lbk \lpar \eps_{{\rm a}}^{2} - \eps_{{\rm b}}^{2} \rpar \lpar \eps_{{\rm a}} \alf_{{\rm a}} -
\eps_{{\rm b}} \alf_{{\rm b}} \rpar \lpar \eps_{{\rm a}} \alf_{{\rm a}} + \eps_{{\rm b}} \alf_{{\rm b}} \rpar^{\ast}
\rbk^{m}}{m'!} \nn
\er
and
\br
{\rm M}_{\gamma}^{(m,m')}(\alf_{{\rm a}},\alf_{{\rm b}};t) &=& L_{m}^{(m'-m)} \lbk - \frac{4 \eps_{{\rm a}} \eps_{{\rm b}}}
{\eps_{{\rm a}}^{2} - \eps_{{\rm b}}^{2}} \frac{\eps_{{\rm b}} \alf_{{\rm a}} + \eps_{{\rm a}} \alf_{{\rm b}}}
{\eps_{{\rm a}} \alf_{{\rm a}} - \eps_{{\rm b}} \alf_{{\rm b}}} \gamma_{{\rm a}} \gamma_{{\rm b}}^{\ast} \rbk F_{m}(t)
F_{m'}^{\ast}(t) + \sqrt{\frac{m+1}{m'+1}} \lpar \eps_{{\rm a}}^{2} - \eps_{{\rm b}}^{2} \rpar \frac{\eps_{{\rm a}}
\alf_{{\rm a}} - \eps_{{\rm b}} \alf_{{\rm b}}}{\eps_{{\rm a}} \alf_{{\rm a}} + \eps_{{\rm b}} \alf_{{\rm b}}} \nn \\
& & \times \, L_{m+1}^{(m'-m)} \lbk - \frac{4 \eps_{{\rm a}} \eps_{{\rm b}}}{\eps_{{\rm a}}^{2} - \eps_{{\rm b}}^{2}}
\frac{\eps_{{\rm b}} \alf_{{\rm a}} + \eps_{{\rm a}} \alf_{{\rm b}}}{\eps_{{\rm a}} \alf_{{\rm a}} - \eps_{{\rm b}}
\alf_{{\rm b}}} \gamma_{{\rm a}} \gamma_{{\rm b}}^{\ast} \rbk G_{m}(t) G_{m'}^{\ast}(t) \; , \nn
\er
with $\gamma_{{\rm a}({\rm b})} = \gamma - \eps_{{\rm a}({\rm b})} \lpar \eps_{{\rm a}({\rm b})} \alf_{{\rm a}} -
\eps_{{\rm b}({\rm a})} \alf_{{\rm b}} \rpar$. In particular, when $\kappa_{{\rm a}({\rm b})} = \kappa$ the expression
for ${\rm K}_{\gamma}(\alf_{{\rm a}},\alf_{{\rm b}};t)$ can be written in the simplified form
\be
\lb{b7}
{\rm K}_{\gamma}(\alf_{{\rm a}},\alf_{{\rm b}};t) = \sum_{m,m'=0}^{\infty} {\rm O}_{\gamma}^{(m)}(\alf_{{\rm a}},
\alf_{{\rm b}}) \lbk {\rm O}_{\gamma}^{(m')}(\alf_{{\rm a}},\alf_{{\rm b}}) \rbk^{\ast} {\cal R}_{\gamma}^{(m,m')}
(\alf_{{\rm a}},\alf_{{\rm b}};t) \; ,
\ee
where
\br
{\rm O}_{\gamma}^{(m)}(\alf_{{\rm a}},\alf_{{\rm b}}) &=& \sqrt{2} \, \exp \lbk - \frac{1}{4} \lpar \left| \alf_{{\rm a}}
+ \alf_{{\rm b}} \right|^{2} + \left| 2 \gamma - \lpar \alf_{{\rm a}} - \alf_{{\rm b}} \rpar \right|^{2} \rpar \rbk
\frac{\lbr \lpar \alf_{{\rm a}} + \alf_{{\rm b}} \rpar \lbk 2 \gamma - \lpar \alf_{{\rm a}} - \alf_{{\rm b}} \rpar
\rbk^{\ast} \rbr^{m}}{2^{m} m!} \; , \nn \\
{\cal R}_{\gamma}^{(m,m')}(\alf_{{\rm a}},\alf_{{\rm b}};t) &=& F_{m}(t) F_{m'}^{\ast}(t) + \half \left| 2 \gamma - \lpar
\alf_{{\rm a}} - \alf_{{\rm b}} \rpar \right|^{2} \frac{G_{m}(t) G_{m'}^{\ast}(t)}{\sqrt{(m+1)(m'+1)}} \; . \nn
\er
This solution is equivalent to consider that the interaction between atom and cavity (external) field has the same
strength. Note that (\ref{b6}) represents an important step in the process of investigation of the effects due the
amplitude of the driving field and the detuning parameter on the nonclassical properties of the cavity field via Wigner
function.

For instance, when the external and cavity fields were described by coherent states, the Wigner function coincides with
${\rm K}_{\gamma}(\alf,\beta;t)$ and for $t=0$, we obtain the initial Wigner function $W_{{\rm a}}^{({\rm c})}(\gamma ; 0)
= 2 \exp (- 2 | \gamma - \alf |^{2})$. Fig. 6 shows the three-dimensional plots of $W_{{\rm a}}^{({\rm c})}(\gamma ; t)$
versus $p = \sqrt{2} \, \ima (\gamma)$ and $q = \sqrt{2} \, \re (\gamma)$ considering the atom-cavity system resonant (a,b)
$\delta = 0$ and nonresonant (c,d) $\delta = 10 \kappa_{{\rm eff}}$ for $| \alf | = 1$ $( \lg {\bf n}_{{\rm a}}
\rg_{{\rm c}} = 1 )$ and $\kappa_{{\rm eff}} t = 100$ fixed, with two different values of amplitude of the driving field:
(a,c) $| \beta | = 2$ $(\lg {\bf n}_{{\rm b}} \rg_{{\rm c}} = 4)$ and (b,d) $| \beta | = 5$ $(\lg {\bf n}_{{\rm b}}
\rg_{{\rm c}} = 25)$. The condition $\kappa_{{\rm a}({\rm b})} = \kappa$ was established in both situations, and the
infinite sums present in (\ref{b7}) were substituted by finite sums as follows:
\bd
{\rm K}_{\gamma}(\alf,\beta;t) = \sum_{m=0}^{\ell} \left| {\rm O}_{\gamma}^{(m)}(\alf,\beta) \right|^{2}
{\cal R}_{\gamma}^{(m,m)}(\alf,\beta;t) + 2 \, \re \lbk \, \sum_{m=0}^{\ell -1} \sum_{m'=m+1}^{\ell} {\rm O}_{\gamma}^{(m)}
(\alf,\beta) \lbk {\rm O}_{\gamma}^{(m')} (\alf,\beta) \rbk^{\ast} {\cal R}_{\gamma}^{(m,m')}(\alf,\beta;t) \rbk \; ,
\ed
where $\ell$ is the maximum value which does guarantee the convergence of this expression (we have fixed $\ell = 50$ in the
numerical investigations). Since the time evolution of composite systems leads us to the essential concept of entanglement
\cite{ref5}, the Figs. 6(a)-(d) reflect the effects of the driving field on the different forms of entanglement in the
tripartite system for a specific value of $\kappa_{{\rm eff}} t$ (maximum entanglement when $\delta = 0$, and minimum
entanglement for $\delta = 10 \kappa_{{\rm eff}}$). For a global view of entanglement of the system under consideration,
different values of $\kappa_{{\rm eff}} t$ and $(|\beta|,\delta)$ are necessary. Here, we give only a partial view of this
important effect.


\end{document}